%% file: RR-6231.tex
\newenvironment{itemize*}%
  {\begin{itemize}%
    \setlength{\itemsep}{0pt}%
    \setlength{\parskip}{0pt}}%
  {\end{itemize}}

\documentclass[a4paper,english]{article}
\usepackage{RR}
\RRNo{6231}
\usepackage{hyperref}
\usepackage{epsfig}
\usepackage{graphics}
\usepackage[english]{babel}
\usepackage{moreverb}

\RRdate{June 2007}
\RRauthor{
Pierre Parrend\and
Stéphane Frénot
}
\authorhead{Parrend \& Frénot}
\RRtitle{Vulnerabilités des Composants Java\\-\\ Une Classification Expérimentale\\Dans le Cadre de la Plate-forme OSGi}
\RRetitle{Java Components Vulnerabilities\\-\\An Experimental Classification \\Targeted at the OSGi Platform   \thanks[sfn]{This Work is partialy founded by Muse IST Project n°026442.}
}

\titlehead{OSGi Vulnerabilities}

\RRresume{La plate-forme d'exécution OSGi rencontre un intérêt grandissant dans deux domaines d'applications différents: les systèmes embarqués, et les serveurs d'applications. Cependant, les propriétés de cette plate-forme relatives à la sécurité ne sont que très peu étudiées, ce qui peut fortement freiner son adoption dans les systèmes industriels. Ceci est d'autant plus critique que la possibilité d'extension dynamique à l'exécution offerte par la plate-forme OSGi rend celle-ci vulnérable à l'injection de code malicieux.

Nous effectuons un audit de l'environnement d'exécution OSGi, afin de créer un catalogue de vulnérabilités. Nous cherchons à référencer les vulnérabilités causées par la spécification `Core', ou bien par la machine virtuelle Java sous-jacente. Les autres éléments définis par OSGi, comme les services standards, ne sont pas considérés.

Afin de mener à bien cet audit, nous définissons un Pattern de Vulnérabilité semi-formel, qui permet de décrire les caractéristiques des vulnérabilités de manière unique, de donner des informations complémentaires, de référencer les protections existantes, et d'identifier le status de l'implémentation des Bundles OSGi de test qui démontrent
chaque vulnérabilité.

A partir de cette analyse, un plate-forme OSGi robuste est construite, et des recommandations pour les spécifications OSGi sont données.
}

\RRabstract{
The OSGi Platform finds a growing interest in two different applications domains: embedded systems, and applications servers. However, the security properties of this platform are hardly studied, which is likely to hinder its use in production systems. This is all the more important that the dynamic aspect of OSGi-based applications, that can be extended at runtime, make them vulnerable to malicious code injection.

We therefore perform a systematic audit of the OSGi platform so as to build a vulnerability catalog that intends to reference OSGi Vulnerabilities originating in the Core Specification, and in behaviors related to the use of the Java language. Implementation of Standard Services are not considered.

To support this audit, a Semi-formal Vulnerability Pattern is defined, that enables to uniquely characterize fundamental properties for each vulnerability, to include verbose description in the pattern, to reference known security protections, and to track the implementation status of the proof-of-concept OSGi Bundles that exploit the vulnerability.

Based on the analysis of the catalog, a robust OSGi Platform is built, and recommendations are made to enhance the OSGi Specifications.
}
\RRmotcle{Plate-forme OSGi$^{tm}$, Sécurité, Java, Plate-forme d'exécution renforcée, Catalogue de Vulnérabilités}
\RRkeyword{OSGi$^{tm}$ Platform, Security, Dependability, Java, Hardened Execution Platform, Vulnerability Catalog}

\RRprojet{ARES}
\RRtheme{\THCom} 

\URRhoneAlpes 

\begin{document}

\makeRR

\tableofcontents{}

\newpage
\listoffigures

\newpage
\listoftables

\newpage
\input{main/introduction}


\newpage
\input{main/classificationBiblio}

\newpage
\input{main/vulnerabilityCharacterization}




\newpage
\input{main/securitySpecifications}

\newpage
\input{main/conclusions}

\newpage
\bibliographystyle{alpha}
\bibliography{biblio_compoV}

\newpage

\appendix
\input{appendix/osgiPlatform}

\newpage
\input{appendix/vulnerabilityLists}

\newpage
\input{appendix/formalVulnerabilityPattern}

\newpage
\input{catalog}

\newpage
\input{appendix/implementations}

\newpage
\input{appendix/xml2tex}

\end{document}

%% file: main/introduction.tex
\section{Introduction}
\label{introduction}

The OSGi Platform, which enables multi-application management over Java Virtual Machines, is currently seeing a dramatic increase in its application domains. First targeted at embedded systems such as multimedia automotive devices, it has since widespread in the world of applications, with the Eclipse Integrated Development Environment, and then to application servers, such as IBM Websphere 6.1, or recent development with JBoss. Sun is envisioning to integrate it in the Sun JVM, and several Java Specification Request (JSR) work groups have been set up on the subject\footnote{http://jcp.org/en/jsr/detail?id=277,http://jcp.org/en/jsr/detail?id=291}.

However, target systems are likely to be highly networked ones, and security implications have so far been mostly overlooked. Actually, the runtime extensibility of applications that is supported by the OSGi platform open a brand new attack vector: code can be installed on the fly, and no mechanism currently guarantees that this code is not malicious. As OSGi-based systems move from Open-Source projects and closed embedded devices toward large-scale systems, this weakness can turn into a major vulnerability, unless security implications are better understood. We therefore perform in this study a systematic analysis of vulnerabilities that are implied by OSGi bundles, and propose adequate counter-measures.

Up to now, Two complementary mechanisms are used to enforce security in the context of OSGi-based systems. The first mechanism is bundle digital signature \cite{osgi05core,parrend06deployment}, which guarantees that only bundles from trusted issuers are installed. This trust requirement forces the issuer to publish only safe bundles, since he will liable for any incident caused by the code he provides. The second mechanism is based on Java permissions, that enable to switch on or off some attack-prone features of the Java Virtual Machine. These mechanisms are mostly insufficient to guarantee that systems are safe. 
First, knowing the identity of a bundle issuer does not give guarantees related to the quality of its bundles. Secondly, most implementations do not have a proper implementation of the digital signature mechanism: they rely on the JVM built-in verification mechanism, which is not compliant with OSGi specifications  \cite{parrend2007sfelix}. And, lastly, Java permissions can not be considered as a panacea, since they are usually not dynamic, and have a great cost in term of functionality, but also in term of performance.

New methods and new security mechanisms therefore need to be defined to provide hardened OSGi Platforms. We present in this report our contribution to this problem, by addressing several requirements. A method for analyzing the security properties of the OSGi Platform is defined. It is based on a catalog of vulnerabilities, and can therefore be completed when further knowledge relative to OSGi Vulnerabilities is gathered. Based on the analysis of this catalog, OSGi specific vulnerabilities are identified, and a prototype is built to show the security mechanisms that can be used. Recommendations for an evolution of the specification of the OSGi Core platform are proposed to enable the OSGi Community to take advantage of this work.

This research report is organized as follows. Works related to vulnerability characterization and analysis are presented in Section \ref{ref:classificationBiblio}. A definition of our Software Vulnerability Pattern is given in Section \ref{sec:vulnerabilityPattern}: it characterized the properties of an OSGi system that need to be listed so as to support vulnerability analysis. The analysis of the vulnerability catalog is then provided, and recommendation for building a hardened OSGi Platform is given in Section \ref{sec:securityRequirements}.

Complementary informations are to be found in the Appendices. In particular, a presentation of the OSGi Platform is given in Appendix \ref{sec:osgiPlatform}; the formal expression of the Vulnerability Pattern we defined is given in Appendix \ref{app:formalVulnerabilityPattern}; the vulnerability catalog in given in its integrality in Appendix \ref{sec:catalog}; and the specific implementations of attacks based on the identified vulnerabilities are given in Appendix \ref{app:implementations}.

%% file: main/classificationBiblio.tex
\section{Characterization of Vulnerabilities in Component-based Systems}
\label{ref:classificationBiblio}

The classification of the security - and vulnerability - properties of systems is necessary to comprehend their weaknesses and to make them more robust. We present here the effort that have been done to establish a precise knowledge of what vulnerabilities are, how to analyse them, and how to take advantage from them to improve the computing systems.

First, the terms that are used to characterize vulnerabilities are defined. Next, the disclosure mechanisms for software flaws are presented. And Vulnerability Patterns that support vulnerability analysis are given.

\subsection{Definitions}
\label{subsec:definitions}

The classification of security properties is based on the distinction between attack, vulnerability, and flaw. The related malicious actions can be prevented by the use of security protections, or countermeasures. The definitions of these terms follow.

\begin{description}
 \item [Security:] the concurrent existence of a) availability for authorized users only, b) confidentiality, and c) integrity \cite{avizienis00dependability}.
 \item [Attacks:] actions that attempt to defeat the expected security status of a system.
 \item [Software vulnerability:] an instance of an error in the specification, development, or configuration of software such that its execution can violate the security policy \cite{Krsul1998softwareVulnerability}.
 \item [Software Flaw:] a flaw is a characteristic of a software system that builds, when put together with other flaws, a vulnerability. The more generic term of WIFF (Weaknesses, Idiosyncrasies, Faults, Flaws) is also used \cite{martin2005enumeration}.
 \item [Security Protection,] or Mitigations, or Countermeasures or Avoidance strategies: mechanisms and techniques that control the access of executing programs to stored information \cite{saltzer73} or to other programs.
\end{description}

Now that the necessary terms are defined, disclosure mechanisms are presented.

\subsection{From Databases to Top-Vulnerability Lists}
\label{subsec:classification}

Vulnerability disclosure aims at providing users and designers informations that enable them to track the security status of their systems. Two main approaches exist: first, vulnerabilities of widespread applications are published in Reference Vulnerability Information (RVI) Databases so as to centralize this information; secondly, these vulnerabilities are classified according to Top-Vulnerability Lists, that support a comprehensive views of potential weaknesses.

\paragraph{Reference Vulnerability Information (RVI) Databases}

Catalogs, Lists and Taxonomies are the favorite vector for expressing the vulnerabilities that are identified in software systems. The approach varies according to the target of the vulnerability identification work. Extensive databases are meant to maintain up to date references on known software vulnerabilities, so as to force the system vendor to patch the error before hackers can exploit it. Taxonomies are particularly used in research works, which foster to improve the knowledge relative to the vulnerabilities. Their goal is to develop tools based on this taxonomies. The drawback of these systematic approaches - catalog and taxonomies - is that they are not easy to remember, and are thus of limited usefulness for developers or code auditor. Several Top Vulnerability Lists have been proposed to solve this problem, and provide the software professionals with convenient practical data.

The main existing references are vulnerability databases. They are also known under the denomination of Refined Vulnerability Information (RVI) sources. Two main types of RVI exists: the vulnerability mailing lists, and the vulnerability databases.

The main mailing lists are the following:

\begin{itemize*}
 \item Bugtraq, 1993 onwards (see http://msgs.securepoint.com/bugtraq/),
 \item Vulnwatch, 2002 onwards (see http://www.vulnwatch.org/),
 \item Full Disclosure, 2002 onwards (see among others http://seclists.org/).
\end{itemize*}

The reference vulnerability databases are the following. They are meant to publish and maintain reference lists of identified vulnerabilities.

\begin{itemize*}
 \item the CERT (Computer Emergency Response Team) Database. It is based on the Common Language for Security Incidents \cite{howard98language}\footnote{http://www.cert.org/}.
 \item the CVE (Common Vulnerabilities and Exposures) Database\footnote{http://cve.mitre.org/}.
 \item the CWE (Common Weaknesses Enumeration) Database. It is bounded with the CWE, and aims at tracking weaknesses and flaws that have not yet turned out to be exploitable for attackers\footnote{http://cwe.mitre.org/index.html}.
 \item the CIAC (Computer Incident Advisory Capability) Database\footnote{http://www.ciac.org/ciac/index.html}.
 \item the OSVDB, Open Source Vulnerability Database\footnote{http://osvdb.org/}. It is centered at Open Source Products.
\end{itemize*}

Complementary Refined Vulnerability Informations Sources are the following organizations: SecuriTeam \footnote{http://www.securiteam.com/}, Packet Storm Security \footnote{http://packetstormsecurity.nl/}, the French Security Incident Response Team \footnote{http://www.frsirt.com/}, ISS X-Force \footnote{http://xforce.iss.net/xforce/alerts}, Secunia, and SecurityFocus.

The limitations of the RVIs is that they follow no stable policy, which makes comparison between sources and between the item of a given sources difficult \cite{Christey2006interpretation}.

\paragraph{Top-Vulnerability Lists}

Since catalogs are not that easy to remember, and therefore to put into practice, several `Top N' lists have been defined. The motivation for such lists is the recurrent drawbacks of other approaches: vulnerability catalogs do not provide a useful overview of the identified vulnerabilities \cite{Christey2006interpretation}.

Therefore, an alternative approach has been developed: to publish lists of prevalent attack categories. Their goal is to be used as reminder for developer or security analysts \cite{mcgraw06softwaresecurity}, and to serve as reference for software product characterization, through integration into security-based code assessment tools \cite{martin2005enumeration}. The most important of these vulnerability lists are presented.

One classification of computer security Intrusions is given by Lindqvist \cite{Lindqvist1997classification} (see appendix \ref{sec:lindqvist}). It contains external and hardware misuse, and several software misuse cases: bypassing intended control, active and passive misuse of resources, preparation for other misuse cases.

The Plover classification\footnote{http://cve.mitre.org/docs/plover/} is an example of rationalization of Vulnerability catalogs to support analysis. It is based on the MITRE CVE Database, and contains 300 specific entries that reflect 1400 vulnerabilities identified in the CVE database. Its goal is to suppress redundancy from the original database, so as to enable scientific analysis,  \emph{e.g.} using statistical approaches \cite{Christey2005plover}.

The Nineteen Deadly Sins of software systems are defined by Michael Howard, from Microsoft \cite{Howard2005_19sins} (see appendix \ref{sec:19}). They describe the most common vulnerabilities that are to be found in enterprise information systems. They concern Web based systems, as well as the architecture of the information systems and the technologies involved.

The Open Web Application Security Project (OWASP) maintains a TOP 10 of Web Applications vulnerabilities\footnote{http://www.owasp.org/index.php/OWASP\_Top\_Ten\_Project} (see the appendix \ref{sec:owasp}). It concerns input validation, data storage, as well as configuration and error management. Another consortium for Web Application security enforcement, the WASC (Web Application Security Consortium), provides its own threat classification\footnote{http://www.webappsec.org/projects/threat/}.)

A convenient vulnerability list is provided by Gary McGraw, through the Seven Kingdoms of software vulnerabilities \cite{mcgraw06softwaresecurity} \cite{tsipenyuk06taxonomy}. The number 7 is chosen to be easily remembered, and each entry is completed with Phyla \emph{i.e.} precise example of the broader categories that are defined by the Kingdoms. The kingdoms are the following: Input Validation and representation, API abuse, Security Features, Time and state, error handling, code quality, encapsulation + environment (see the appendix \ref{sec:7}). This classification is targeted at enterprise information systems.

\medskip

The publication of newly discovered vulnerabilities and of Top-Lists helps the practitioner stay informed of the actual security risks of the system they use, but they provide little support for systematic analysis. Vulnerability Patterns must be defined to formalize vulnerability informations.

\subsection{Vulnerability Patterns}
\label{subsec:vul_patterns}

The descriptive spirit of Design Pattern \cite{Alexander1977patternlanguage}, \cite{gamma94pattern}, \cite{Mowbray1997corbaDP},  is well suited for application in the security fields, where the question of organization and exploitation of the knowledge is central to the protection of systems - and not straightforward, if one judges from the various approaches that are used. Two types of patterns are defined in the security domain: Attack Patterns, and Vulnerability Patterns.

Attack Patterns represent potential attacks against a system. They model the preconditions, process and postconditions of the attack. They can be combined with attack trees, so as to automate the identification of attacks that are actually build from simpler atomic attacks \cite{moore01modeling}. An extensive presentation of the application of attack pattern is given in the book by Markus Schumacher \cite{Schumacher2003}. The use of Attack Patterns together with with software architecture description to identify vulnerabilities is described by Gegick \cite{Gegick2005attackPatterns}.
The limitation of this approach is that the attacks must be modelized, but the system must also be, which makes this approach impractical, and often not realistic based on the actual knowledge that is available on systems.

The Vulnerability Patterns are used in the catalog of vulnerabilities. They often contain a limited number of information that are meant to identify the vulnerability, but also to not make it easily reproduceable without a reasonable amount of effort, to prevent lazy hackers to exploit the vulnerability databases as a source of ready-to-exploit attack references.

We list here the most wide-spread Vulnerability Patterns, along with the attribute they contain:

\begin{itemize*}
 \item Rocky Heckman pattern\footnote{http://www.rockyh.net/}: Name, type, subtype, AKA, description, more information;
 \item CERT (Computer Emergency Response Team) pattern:  name, date, source, systems affected, overview, description, qualitative impact, solution, references;
 \item CVE\footnote{http://cve.mitre.org/} (Common Vulnerability and Exposures) pattern: name, description, status, reference(s);
 \item CIAC\footnote{http://www.ciac.org/ciac/index.html} (US Department of Energy) pattern: identifier, name, problem description, platform, damage, solution, vulnerability assessment, references.
\end{itemize*}

These Vulnerability Patterns are quite simple ones. They have an informative goal, but do not intend as other patterns do at supporting the reproduction of the vulnerability with a minimum of effort. This approach makes sense relative to their use context - making users and administrators aware of the existence of the flaws - but are not sufficient to support detailed analysis of the related vulnerabilities.

So as to support the automation of the security  process, and to make vulnerability analysis possible, it is necessary to put constraints on the Vulnerability Patterns. This is performed through the definition of taxonomies, which provide a fine grain description of the properties of each vulnerability.

Each taxonomy should verify the properties of a valid taxonomy, as defined by \cite{Krsul1998softwareVulnerability} and \cite{howard98language}. These properties are the following: objectivity, determinism, repeatability, specificity (disjunction), observability.

The seminal work on vulnerability taxonomy has been performed by Abbott \cite{Abbott1975os} and Bisbey \cite{Bisbey1978protectionAnalysis}. The flaws are classified by type of error (such as incomplete Parameter validation). This approach turns out not to support deterministic decisions, since one flaw can often be classified in several categories according to the context. To solve this problem, Landwehr \cite{landwehr94taxonomy} defines three fundamental types of taxonomies for vulnerabilities: classification by genesis of the vulnerability, by time of introduction, and by location (or source). 

Moreover, vulnerabilities should be considered according to specific constraints or assumptions, since there existence most of the time depends on the properties of the environment \cite{Krsul1998softwareVulnerability}. This assumptions make it necessary to rely on a well defined system model. For generic computing systems, such a model is proposed by the Process/Object Model \cite{Bazaz2006contraints}. This requirement makes it impossible for generic purpose databases to rely on specific taxonomies. For instance, the Common Vulnerability Enumeration \cite{Baker1999cve} project has given up the use of taxonomies. An explicit system model must thus be available to support vulnerability taxonomies, and therefore the possibility of security automation or analysis.

Extensive discussions of vulnerability taxonomies can be found in \cite{Krsul1998softwareVulnerability}, \cite{Seacord2005vulnerabilities}, \cite{Weber2005softwareflaws}. The CWE (Common Weaknesses Enumeration) Project maintains a web page with additional references, and a graphical representation of each taxonomy\footnote{http://cwe.mitre.org/about/sources.html}.

\bigskip

In this section, fundamental concepts of vulnerability analysis have been introduced: definitions have been given to provide a firm basis to work on, and the existing works in the domain of vulnerability analysis have been presented. This work concerns Vulnerability properties, which are often presented under the form of a taxonomy, and Vulnerability Patterns, which gather the information concerning several properties in a formalized way.

 Existing Properties and Patterns are not sufficient to describe the vulnerabilities of an OSGi Platform, for several reason: first, they do not take explicitly into account the presence of a virtual machine; secondly, they are usually targeted at monolithic systems, whereas OSGi provides a high degree of modularity through the bundles and the dependency resolution. We therefore first need to define the properties of interest for an OSGi-based System, as well as a suitable Pattern, before the actual vulnerabilities of the platform can be analyzed.

%% file: main/vulnerabilityCharacterization.tex
\section{The Semi-formal Software Vulnerability Pattern}
\label{sec:vulnerabilityPattern}

The goal of this study is to identify and to characterize the vulnerabilities of the OSGi platform, which is introduced in the Appendix \ref{sec:osgiPlatform}. This characterization is to be done with a set of specific properties, and organized in a semi-formal Vulnerability Pattern. Existing references are not sufficient to describe the vulnerabilities of the OSGi Platform: neither virtualization nor componentization, that are provided in the context of OSGi, are taken into account. Moreover, we want our Vulnerability Pattern to provide us with enough information to patch them or build suitable security mechanism, which is not the case in the literature.

The properties of interested that are tracked are taken from existing software security taxonomies. We add a new entry, the `Consequence Description', that aims at evaluating the seriousness of the vulnerability. The Pattern is made up of four parts:
\begin{itemize*}
 \item a Reference, for rapid consultation,
 \item a Description part, for additional and potentially more verbose information, 
 \item an Implementation part, to identify the test conditions of the vulnerability, 
 \item a Protection part, because the objective of identifying the vulnerability is to be able to patch them.
\end{itemize*}

Our experimental process is the following. First, known flaws that can affect Java code \cite{bloch01effectivejava, bloch05puzzlers} have been identified, and their impact on an OSGi Platform has been tested. Secondly, potentially dangerous mechanisms, such as native code execution, have been selected from related projects. The third source of information in our quest for vulnerabilities of OSGi bundles is the specifications of the elements that make up an OSGi platform: the Java Virtual Machine, and the OSGi platform itself. Several Java API let the code access to the Virtual Machine itself (\emph{e.g.} the System or Runtime API), or are known to cause the execution hang (Threads). The behavior of the OSGi platform in the presence of malformed or malicious bundles is not specified. We therefore review the various entities that make up this execution environment: the format of the bundle meta-data (Manifest File), the registration of services, the bundle management and fragment functionalities. For each potential vulnerability, we implemented a malicious bundle. This makes possible to validate the hypothesis, and to identify the conditions for each attack. When protections against these attacks exist, they are validated through experiment. The attack bundles are tested in the four main Open Source implementations of OSGi, Felix, Knopflerfish, and Equinox, and Concierge.

We focus on the behavior of the core of the considered execution environment, which comprises the JVM and the OSGi platform. We therefore do not consider the management tools for Java systems, such as JMX, or JVM TI. JMX enables to manage a JVM though code executed inside it. JVM TI is a C library that makes full control over the JVM possible through a third party program, which can then access the available threads, provides an extensive debugging of the platform, and control the JVM state. 
Secondly, the OSGi bundles communicate through services they publish inside the framework. According to the type of data they handle, these services can be subject to specific vulnerabilities.  A list of Service-Level vulnerabilities is given by the Findbugs reference list (http://findbugs.sourceforge.net/bugDescriptions.html, `Malicious Code Vulnerability' category).
Lastly, the OSGi specification defines a bunch of standard services (HTTP, device, service wiring, UPnP services, etc.). We do not consider these services either.

A Vulnerability Pattern is defined to normalize the information gathered relative to each vulnerability (see section \ref{subsec:vulnerabilityPattern}). The taxonomies for each properties of interest are given and explained in section \ref{subsec:characterization}. An example is given in section \ref{subsec:anExample} to highlight the information provided by the Vulnerability Pattern.

\subsection{The Structure of the Semi-formal Vulnerability Pattern}
\label{subsec:vulnerabilityPattern}

The characteristics of interest to describe the vulnerabilities of a software system need to be gathered in a coherent set that contains all the informations that are useful to understand and prevent these vulnerabilities. We therefore define a `Semi-formal Vulnerability Pattern' that is similar to the  `Attack Patterns' \cite{moore01modeling}. On the opposite of this latter, the Vulnerability Pattern is centered around the identified vulnerability, so as to make their correction easy. Existing vulnerability patterns, which are presented in the section \ref{ref:classificationBiblio}, can not be reused as-is, since they provide not enough details for our purpose.

The Vulnerability Pattern is compound of following informations. Its formal expression is given in the Apendix \ref{app:formalVulnerabilityPattern}.

\medskip

\begin{tabular}{|l|} 
\hline
\begin{minipage}{400pt}

\bigskip

\paragraph{Vulnerability Reference}

\begin{itemize*}
  \item \textbf{Vulnerability Name:} The descriptive name of the vulnerability
  \item \textbf{Identifier:} a unique identifier for each vulnerability. In our catalog, the identifier is built out of the catalog identifier, the abbreviation of the source entity, and the number ID of the vulnerability in the catalog for the related source entity.
  \item \textbf{Origin:} The bibliographic reference of the vulnerability.
  \item \textbf{Location of Exploit Code:} Where the code that performs the attack is located in the malicious Bundle (see Figure \ref{fig:location}).
  \item \textbf{Source:} the entity in the execution platform that is the source of the vulnerability, along with the exact flaw or functionality causing it.
  \item \textbf{Target:} the target of the attack that can be performed through the vulnerability, \emph{i.e.} the victim of the attack  (see figure \ref{fig:attack-targets}).
  \item \textbf{Consequence Type:} the type of consequence of an attack exploiting this vulnerability (see figure \ref{fig:consequences}).
  \item \textbf{Time of Introduction:} the Life Cycle phase where the vulnerability is introduced. Corrective measures can be taken at this time. Security measures can be taken in subsequent phases so as to prevent the exploitation of the vulnerability  (see figure \ref{fig:introduction-time}).
  \item \textbf{Time of Exploit:} the life-cycle phase where the vulnerability can be exploited  (see figure \ref{fig:vulnerability-exploit-time}).
This is the last phase where security measures can be undertaken.
\end{itemize*}

\medskip

\end{minipage}
\\ \hline
\end{tabular}

\begin{tabular}{|l|} 
\hline
\begin{minipage}{400pt}

\bigskip

\paragraph{Vulnerability Description}

\begin{itemize*}
  \item \textbf{Description:} a description of the attack
  \item \textbf{Preconditions:} properties if the systems that must be true so as to make the exploitation of the vulnerability possible.
  \item \textbf{Attack Process:} description of the process of exploitation of the vulnerability.
  \item \textbf{Consequence Description:} more information relative to the consequences of an attack using this vulnerability.
  \item \textbf{See also:} other vulnerabilities based on similar attack sources.
\end{itemize*}

\paragraph{Vulnerability Implementation}

\begin{itemize*}
  \item \textbf{Code Reference:} the reference of the implementation code (\emph{i.e.} the name of the malicious OSGi bundle.)
  \item \textbf{Concerned OSGi Profile:} the OSGi profile(s) where this vulnerability exists.
  \item \textbf{Date:} the date of the creation of the Vulnerability Pattern (for reference)
  \item \textbf{Test Coverage:} the percentage of the known implementations of the vulnerability that have been implemented in a test bundle. The identified implementations for the main attacks are given in the Appendix \ref{app:implementations}.
  \item \textbf{Tested on:} the OSGi Platform Implementations for which this vulnerability have been tested.
\end{itemize*}

\paragraph{Protection}

\begin{itemize*}
  \item \textbf{Existing mechanisms:} available protections to prevent this vulnerability from being exploited.
  \item \textbf{Life-cycle enforcement point:} the life-cycle phase where the protection mechanisms must be enforced.
  \item \textbf{Potential mechanisms:} protections that could be developed so as to prevent this vulnerability from being exploited.
  \item \textbf{Attack Prevention:} the measures that can be taken to prevent an attack based on this vulnerability to be fulfilled, even if it is launched.
  \item \textbf{Reaction:} the correction action that can be taken to recover from a successful attack.
\end{itemize*}

\medskip

\end{minipage}
\\ \hline
\end{tabular}

\subsection{Vulnerability Taxonomies for OSGi-based Systems}
\label{subsec:characterization}

Before analyzing each vulnerability, it is necessary to identify the properties of interest that need to be characterized. Moreover, the potential values for each property should be identified, and build a properly defined taxonomy. Following aspects need to be considered: the reference of the attack bundle implementation that takes advantage of the vulnerability, the life-cycle characteristics of the vulnerability, so as to know when the vulnerability is introduced, and when it is exploited, and the existing and potential security mechanisms.

The properties we selected to be included in the vulnerability pattern are thus the following:

\begin{itemize*}
 \item the reference of the vulnerability pattern (to identify the code, and the condition of the experiments),
 \item the location of the malicious code,
 \item the source of the flaw(s) in the system (and the specific flaw(s) and/or the dangerous functionality(ies)),
 \item the target of attacks based on the vulnerability,
 \item the consequences of the related attack, and
 \item the time of introduction of the vulnerability,
 \item the time of exploitation of the vulnerability,
 \item the exiting protections against this attack,
 \item the potential protections against this attack.
\end{itemize*}

The goal of these properties is to make the information explanatory, predictive \cite{Krsul1998softwareVulnerability}, but also useful \cite{Weber2005softwareflaws}. Explanatory, because the vulnerability should be intuitively understood by the persons who consult the vulnerability catalog we propose, even with little previous knowledge of the OSGi Platform. Predictive, because the potential values of each characteristic should cover the whole field of possible options, or to explain why some are not concerned. Useful, because the objective of a vulnerability catalog is to highlight the security requirements of the platform under study. The conclusions of the analysis is presented in the section \ref{sec:securityRequirements}.

For each property, a taxonomy is defined, that contains the values this property can take. Two approaches are used to define this taxonomy. It is either defined a priori, \emph{i.e.} before the catalog is completed, or a posteriori, with data that are identified during the experiments.


We now present the taxonomy for each of the properties of interest.

The first two properties of interest are the Location of the malicious code and the source of the vulnerability. The location concerns the place in the attack bundle where the attack `payload' is located. The source indicates which entity in the execution environment is responsible for the vulnerability, \emph{i.e.} for the system behavior that opens the door to the attack.

\begin{figure}[htb]
\centering
\epsfig{file=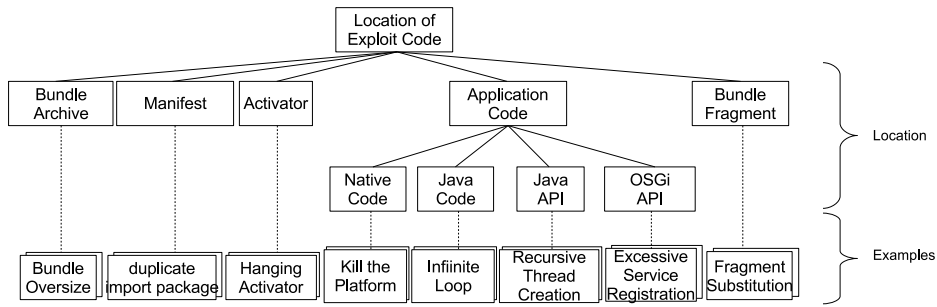, width=350pt}
\caption{Potential Locations of malicious Code in a Bundle}
\label{fig:location}
\end{figure}

Figure \ref{fig:location} shows the potential locations of malicious code inside a malevolent bundle. The malicious code can be located in the archive structure (such as archive oversize, or a decompression bomb), in the manifest (such as duplicate imports, which make the installation abord), or in the bundle Activator (is this latter is hanging). It can also be located in the applicative code of the bundle, being native code, Java code, the Java APIs, or the OSGi API. The malicious code can also be located in fragment, which are specific bundle types.

\begin{figure}[htb]
\centering
\epsfig{file=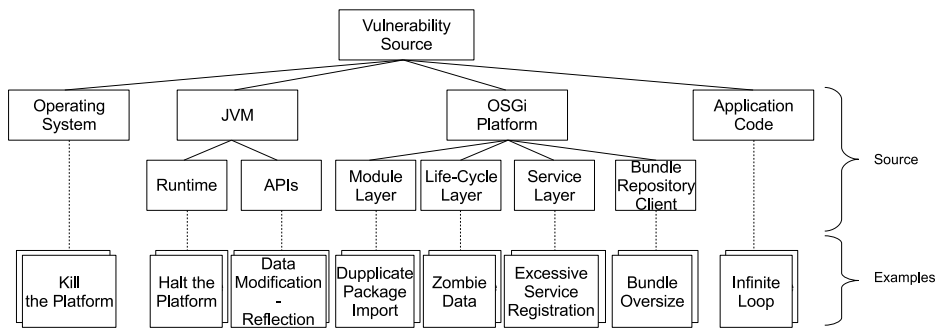, width=350pt}
\caption{Vulnerability Sources in an OSGi-based System}
\label{fig:vulnerability-sources}
\end{figure}

The actual sources of vulnerabilities match the different Layers that are defined by the OSGi Specification, along with the Bundle Repository client which enables installation from remote bundles, and various code properties such as Services, the JVM APIs, or the algorithmic properties of the programs. They are shown in Figure \ref{fig:vulnerability-sources}.

\begin{figure}[htb]
\centering
\epsfig{file=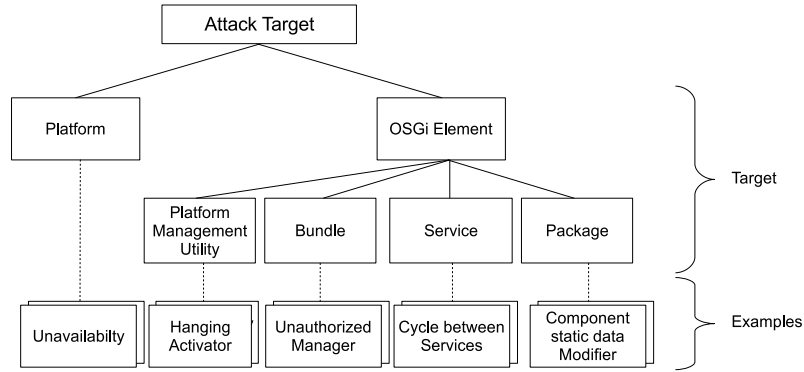, width=300pt}
\caption{Potential Targets of Attacks against an OSGi Platform}
\label{fig:attack-targets}
\end{figure}

Figure \ref{fig:attack-targets} shows the potential targets of attacks against an OSGi Platform. These targets can be either the whole platform, or specific OSGi Elements. Attacks against the whole Platform can for instance result in complete unavailability if this latter. The victim OSGi Elements can be the Platform Management Utility, which makes it possible for the user to control the life-cycle of bundles (the activator can hang), the bundle itself (which can be started or stopped), Services (which can suffer from cycles) or packages (for instance, static data that is by default not accessible could be modified through Bundle Fragments).

\begin{figure}[htb]
\centering
\epsfig{file=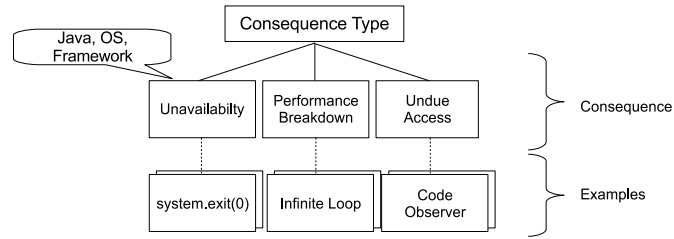, width=250pt}
\caption{Consequences of the Vulnerabilities of the OSGi Platform}
\label{fig:consequences}
\end{figure}

Figure \ref{fig:consequences} shows the potential consequences of an attack against an OSGi Platform. Three types of consequences are identified: Unavailability, Performance Breakdown, and Undue Access. Unavailability can be cause by stopping the platform; Performance Breakdown can be the result of an infinite loop; and Undue Access can be performed through Fragments or through Reflection over Services.

\begin{figure}[htb]
\centering
\epsfig{file=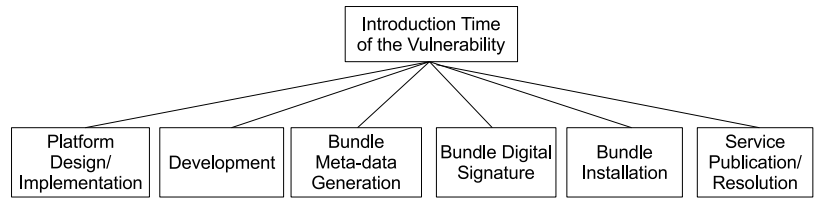, width=300pt}
\caption{Introduction Time for the identified flaws}
\label{fig:introduction-time}
\end{figure}

Figure \ref{fig:introduction-time} shows the actual introduction time of the vulnerabilities. The introduction time can be as early as the design and implementation of the platform (when the flaw originates in the platform), or be the development time, the generation of the Meta-data of the bundles, the digital signature of the bundle,  the installation, or even the publication and resolution of services.

\begin{figure}[htb]
\centering
\epsfig{file=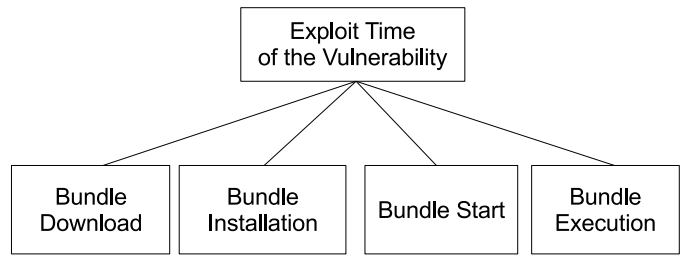, width=250pt}
\caption{Exploit Time for the identified Flaws}
\label{fig:vulnerability-exploit-time}
\end{figure}

The taxonomy for Time of Exploit of the vulnerability is represented in figure \ref{fig:vulnerability-exploit-time}.
 This time of exploit concerns necessarily the Life-Cycle steps inside the execution platform. They can therefore be: the download, installation, Bundle start (if the vulnerability is present in the bundle activator) or execution time (either through service call, or through use of exported packages).

The existing protections against attacks based on the identified vulnerability are the following: only runtime execution permissions, either at the JVM level or at the OSGi Platform level, are currently available to protect an OSGi Platform from hackers. We propose our own implementation of the OSGi Bundle Digital Signature validation process, which is part of the OSGi Security Layer.

\begin{itemize*}
  \item Java Permissions,
  \item OSGi Permissions (in particular AdminPermission),
  \item SFelix implementation of the Bundle Digital Signature Validation.
\end{itemize*}

The properties of interest that characterize a vulnerability have been presented. Next paragraph gives the full Vulnerability Pattern that is based on these properties, and adapted for better comprehension.

\subsection{A Vulnerability Example: `Management Utility Freezing - Infinite Loop'}
\label{subsec:anExample}

So as to highlight the role of the defined Vulnerability Pattern, we now present an example of vulnerability; the `Management Utility Freezing - Infinite Loop' vulnerability. The whole vulnerability catalog is given in the Appendix \ref{sec:catalog}.
This vulnerability consists in the presence of an infinite Loop in the activator of a given Bundle, which causes the platform management tool (often an OSGi shell) to freeze. The presence of infinite loops as a vulnerability is given by Blooch is `Java Puzzlers - Traps, Pitfalls and Corner Cases', puzzlers 26 to 33 \cite{bloch05puzzlers}. The matching Pattern is first given, and then explained.

\paragraph{The Vulnerability Pattern}

\input{vulnerabilityPatterns/managementUtilityFreezingInfiniteLoop2}

\paragraph{Details of the Vulnerability Pattern}

The `Management Utility Freezing - Infinite Loop' is referenced under the identifier `mb.osgi.4', which means `malicious bundle catalog - originates in the OSGi Platform itself - number 4. This vulnerability is an extension of the `Infinite Loop in Method Call' one. It has been identified in the frame of the research project `Malicious Bundles' of the INRIA Ares Team..

The location of the malicious code that performs the attack is the Bundle Activator. Its source is the Life-Cycle Layer of the OSGi Platform, which is not robust against such a vulnerability. Its target is the Platform Management Utility, which can be either the OSGi shell or a graphical interface such as the Knopflerfish GUI. This vulnerability has a two-fold consequence: the method does not return, so that the caller also freezes; and the infinite loop consumes most of the available CPU, which causes the existing services to suffer from a serious performance breakdown.

This vulnerability is introduced during development, and exploited at bundle start time.

Related Vulnerability Patterns are `Management Utility Freezing - Hanging Thread', that also targets the Management Utility, `Infinite Loop in Method Call', `CPU Load Injection', `Stand-alone Infinite Loop' that have the same consequence of performance breakdown, and the `Hanging Thread', that also freezes the calling thread.

No specific protection currently exists. Two potential solutions have been identified. The first consists in launching every Bundle Activator in a new Thread, so as not to block the caller if the activator hangs. The second solution would enable to prevent invalid algorithms to be executed: static code analysis techniques such as Proof Carrying Code or similar approaches \cite{necula97proofcarrying} can provide formal proves of code wellformedness.

Its reference implementation is available in the OSGi bundle named `fr.inria.ares.infinite- loopinmethodcall-0.1.jar', referenced the 2006-08-24. The test coverage is 10 \%, since ten types of infinite loops have been identified (see the appendix \ref{app:implementations}), and only one has been implemented. The test bundle have been tested on the following implementations of the OSGi platform: Felix, Equinox, Knopflerfish, and Concierge. The only robust Platform is our SFelix Platform, which is a prototype meant to enhance to current Felix implementation.

\bigskip

This example highlights the informations that can be found in each Vulnerability Patterns. The information related to the other vulnerabilities is given under the form of patterns, to provide a quick overview of the characteristics, and to make analysis possible. The catalog of the vulnerability patterns is presented in the section \ref{sec:catalog}. The section \ref{sec:securityRequirements} presents the analysis of this catalog, and the security requirements can can be deduced from it.

%% file: vulnerabilityPatterns/managementUtilityFreezingInfiniteLoop2.tex
\paragraph{Vulnerability Reference}

\begin{itemize*}
  \item \textbf{Vulnerability Name:} Management Utility Freezing - Infinite Loop
  \item \textbf{Extends:} Infinite Loop in Method Call
  \item \textbf{Identifier:} Mb.osgi.4
  \item \textbf{Origin:} Ares research project `malicious-bundle'
  \item \textbf{Location of Exploit Code:} Bundle Activator
  \item \textbf{Source:} OSGi Platform - Life-Cycle Layer (No safe Bundle Start)
  \item \textbf{Target:} OSGi Element - Platform Management Utility
  \item \textbf{Consequence Type:} Performance Breakdown;
	Unavailability
  \item \textbf{Introduction Time:} Development
  \item \textbf{Exploit Time:} Bundle Start
\end{itemize*}

\paragraph{Vulnerability Description}

\begin{itemize*}
  \item \textbf{Description:} An infinite loop is executed in the Bundle Activator
  \item \textbf{Preconditions:} -
  \item \textbf{Attack Process:} An infinite loop is executed in the Bundle Activator
  \item \textbf{Consequence Description:} Block the OSGi Management entity (the felix, equinox or knopflerfish shell; when launched in the KF graphical interface, the shell remain available but the GUI is frozen). Because of the infinite loop, most CPU resource is consumed
  \item \textbf{See Also:} CPU Load Injection, Infinite Loop in Method Call, Stand Alone Infinite Loop, Hanging Thread
\end{itemize*}

\paragraph{Protection}

\begin{itemize*}
  \item \textbf{Existing Mechanisms:} -
  \item \textbf{Enforcement Point:} -
  \item \textbf{Potential Mechanisms:} Code static Analysis ;
	Resource Control and Isolation - CPU ;
	OSGi Platform Modification - Bundle Startup Process (launch the bundle activator in a separate thread to prevent startup hanging)
  \item \textbf{Attack Prevention:} -
  \item \textbf{Reaction:} -
\end{itemize*}

\paragraph{Vulnerability Implementation}

\begin{itemize*}
  \item \textbf{Code Reference:} Fr.inria.ares.infiniteloopinmethodcall-0.1.jar
  \item \textbf{OSGi Profile:} J2SE-1.5
  \item \textbf{Date:} 2006-08-24
  \item \textbf{Test Coverage:} 10\%
  \item \textbf{Known Vulnerable Platforms:} Felix;
	Equinox;
	Knopflerfish;
	Concierge
  \item \textbf{Known Robust Platforms:} SFelix
\end{itemize*}

%% file: main/securitySpecifications.tex
\section{Requirements for secure OSGi Systems}
\label{sec:securityRequirements}

The analysis of the Vulnerability Patterns we presented in the catalog provides guidelines for programming secure OSGi-based systems. The objective here is two fold. First, the weaknesses of the OSGi Platform are to be identified, so as to provide a framework for the evolution of its specification. Secondly, these weaknesses are to be made available in a developer-compliant way, so that programmers can refer to them to verify that their system do not open the way to known attacks: the \emph{Seven Deadly Sins} of the OSGi R4 Platform are therefore defined.

 These guidelines are - off course - based on the catalog at the moment of its publication, and can therefore evolve in the future, when new vulnerabilities will be discovered, or when new part of OSGi systems will be considered. Actually, the management tools such as JVMTI, and the OSGi standard services are not considered, and can be the source for new vulnerabilities.

This section is organized as follows. Subsection \ref{subsec:analysis} presents the analysis of the catalog through statistics related to the significant properties of the vulnerabilities. Subsection \ref{subsec:hardenedOSGi} presents the Security Requirements for a hardened OSGi Platform. Lastly, Subsection \ref{subsec:specifications} gives a series of recommendation for the OSGi Specification, in order to make the platform more robust.

\subsection{Catalog Analysis}
\label{subsec:analysis}

The analysis of the identified Vulnerability Patterns provides quantitative data relative to these vulnerabilities. This subsection provides a summary of the significant properties that characterize a vulnerability in an OSGi Execution Environment. We use the term \textbf{OSGi Execution Environment} to describe an execution platform running OSGi on top of a JVM. This denomination highlights the fact that not all vulnerabilities are bound to the OSGi specification itself, but can also originate in other parts of the system.

First, quantitative results relative to the vulnerability sources, functions, flaws, as well as the identified attack targets are given. Next, a summary of the vulnerabilities for each tested OSGi Platform implementation is presented.

\begin{figure}[htb]
\centering
\epsfig{file=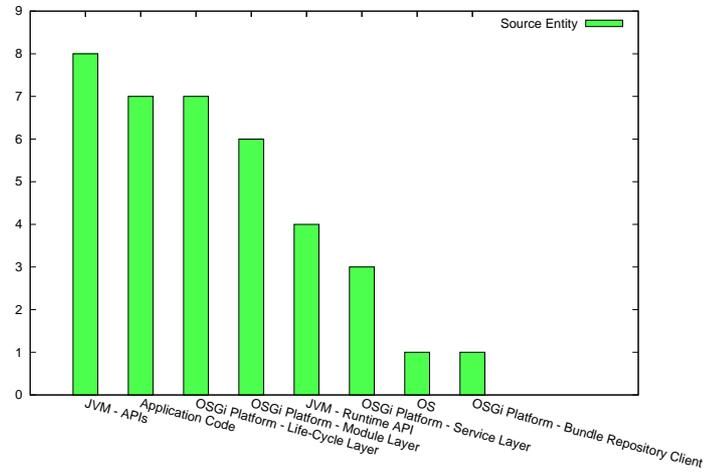, width=250pt}
\caption{Entities that are Source of the vulnerabilities}
\label{fig:source}
\end{figure}

Figure \ref{fig:source} shows the source entity of the identified vulnerabilities. The most important source is the Java API, which causes the bigger part of the vulnerabilities. Next, the Application Code properties, the OSGi Life-Cycle Layer and the OSGi Module Layer also generate an important number of vulnerabilities. Next comes the Java Runtime API, which is particularly sensitive, and the OSGi Service Layer. The Operating System and the Bundle Repository Client must also be considered as potential source of vulnerabilities, even though their impact is more marginal.

\begin{figure}[htb]
\centering
\epsfig{file=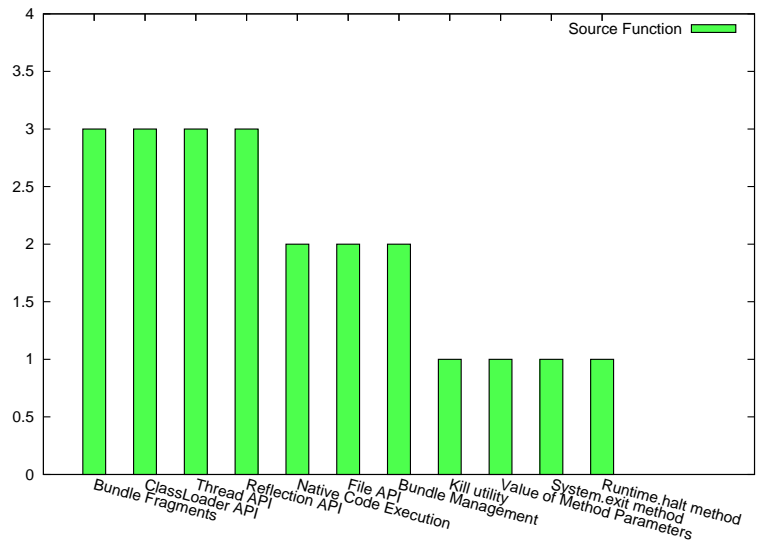, width=300pt}
\caption{Functions that prove to be dangerous in the context of an OSGi Platform}
\label{fig:functions}
\end{figure}

Figure \ref{fig:functions} shows the cardinality of the identified dangerous functions. First come the OSGi Bundle Facility, and the Java APIs Reflection, ClassLoader, and Thread. Next, the bundle management, the Java File API and the opportunity of executing native code open the way to abuses. Several other functions prove to be dangerous: the Runtime.halt() and the System.exit() methods, the lack of control on method parameters, and the kill utility at the OS level which can be used to shut the platform down.

\begin{figure}[htb]
\centering
\epsfig{file=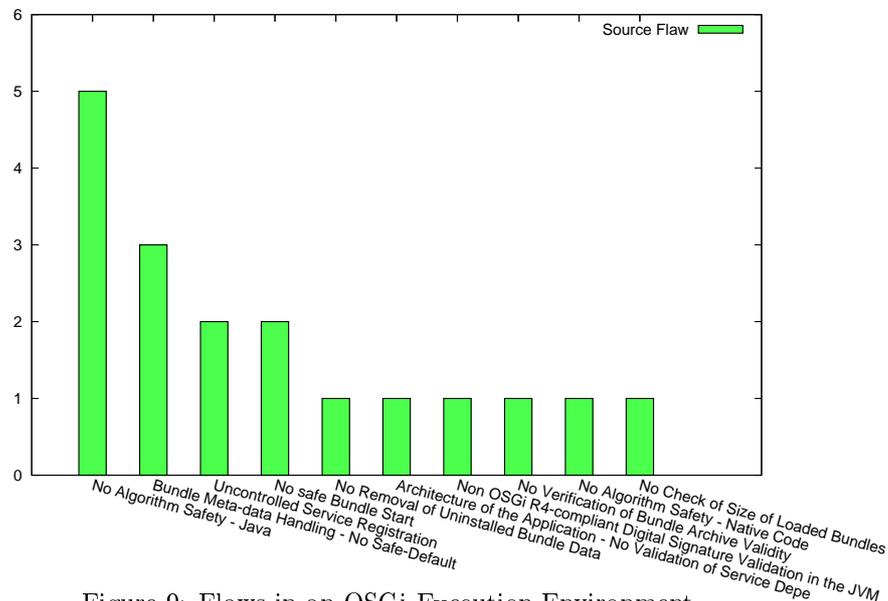, width=300pt}
\caption{Flaws in an OSGi Execution Environment}
\label{fig:flaws-card}
\end{figure}

Figure \ref{fig:flaws-card} shows the cardinality of the flaws of a Java-based execution environment with the OSGi Platform. The most important flaw is the lack of algorithm safety in the Java language. Next come several properties of the OSGi platforms, such as the lack of safe-default bundle meta-data handling during the dependency resolution phase, the lack of control on the service registration process, and the lack of robustness of the bundle start mechanism, which heavily relies on the validity of the bundle activators. Several other punctual flaws have been identified: data of uninstalled bundles is often kept on the disk space, being not accessible, no dependency control is performed at the service level, the process of Digital Signature validation is sometimes not compliant with the OSGi R4 Specifications, and the bundle archive is never checked for size or validity, which provides no protection against decompression bombs or large files, in particular in resource-constraint environments.

\begin{figure}[htb]
\centering
\epsfig{file=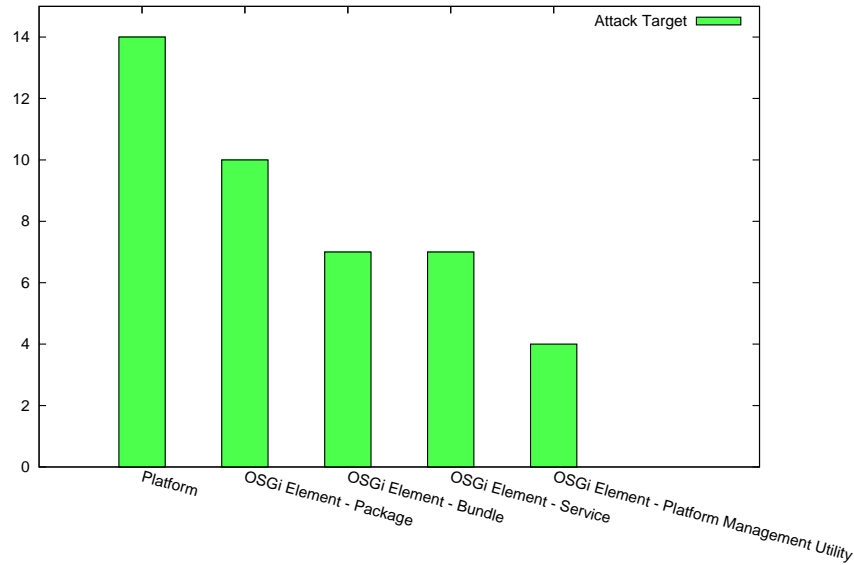, width=300pt}
\caption{Targets of Attacks against an OSGi Execution environment}
\label{fig:targets}
\end{figure}

Figure \ref{fig:targets} shows the target of attacks against an OSGi execution environment. The entity that is the first target of the identified attacks is the platform. This means that most of the identified attack can easily prevent all services on the platform the be executed in a satisfactorily manner. OSGi specific elements, such as packages, Bundles or services are other frequent targets. Lastly, the Platform Management Utility can also be targeted, which would not prevent the platform to provide existing services, but would prevent any evolution of these services - as well as the removal of the malicious bundle.

The table \ref{tab:platform_vulnerabilities} shows a summary of the properties of the OSGi Platform implementations under study. The considered platforms are the main Open Source Projects: Felix\footnote{http://cwiki.apache.org/FELIX/index.html}, Knopflerfish\footnote{http://www.knopflerfish.org/}, Eclipse Equinox\footnote{http://www.eclipse.org/equinox/}, and Concierge\footnote{http://concierge.sourceforge.net/}. For comparison, we also provide the data related to SFelix, which is the Hardened OSGi Platform we develop. It is based on the Felix Platform 0.8.0.

Most Open Source OSGi Platforms are very fragile regarding the set of vulnerability we identified. Equinox proves to have slightly better results than the other ones. SFelix currently does not intend to provide protections against all the identified vulnerabilities, but only to provide a first enhancement of current implementations.

\input{tables/PlatformVulnerabilities}

The result of our analysis have been presented for the properties that characterize vulnerabilities of an OSGi Platform: Location of the malicious payload in Bundles, Vulnerability Source, Flaws and dangerous Functions, as well as the identified Attack Targets. It is now possible to identify the requirement for a Hardened OSGi Platform.

\subsection{Requirements for a Hardened OSGi Platform}
\label{subsec:hardenedOSGi}

The requirement for a Hardened OSGi Platform can be deduced from the actual and potential security mechanisms. The objective is to highlight the security mechanisms that need to be better exploited (for the existing ones), and the ones that need to be developed (for the potential ones). Priorities can be set according to the type of target and consequences of the attacks: it is in any case worth preventing an attack that makes the whole platform unavailable, but it may be less important to prevent attacks that provoke only the unavailability of the malicious bundle itself.

\begin{figure}[htb]
\centering
\epsfig{file=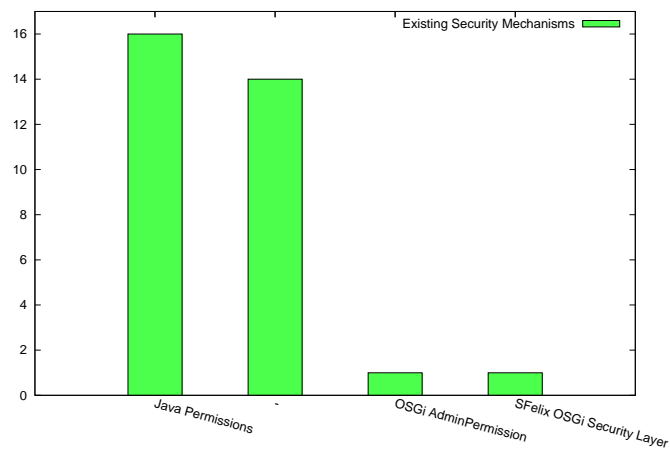, width=250pt}
\caption{Actual Protection Mechanisms}
\label{fig:existing_mechanisms}
\end{figure}

Figure \ref{fig:existing_mechanisms} shows the cardinality of the actual protection mechanisms. Most of the vulnerabilities can be prevented by Java Permissions. However, an important number of them currently have no associated protections. The OSGi Admin Permission and the SFelix implementation of the OSGi Security Layer (Digital Signature Validation part) account each for one vulnerability.

\begin{figure}[htb]
\centering
\epsfig{file=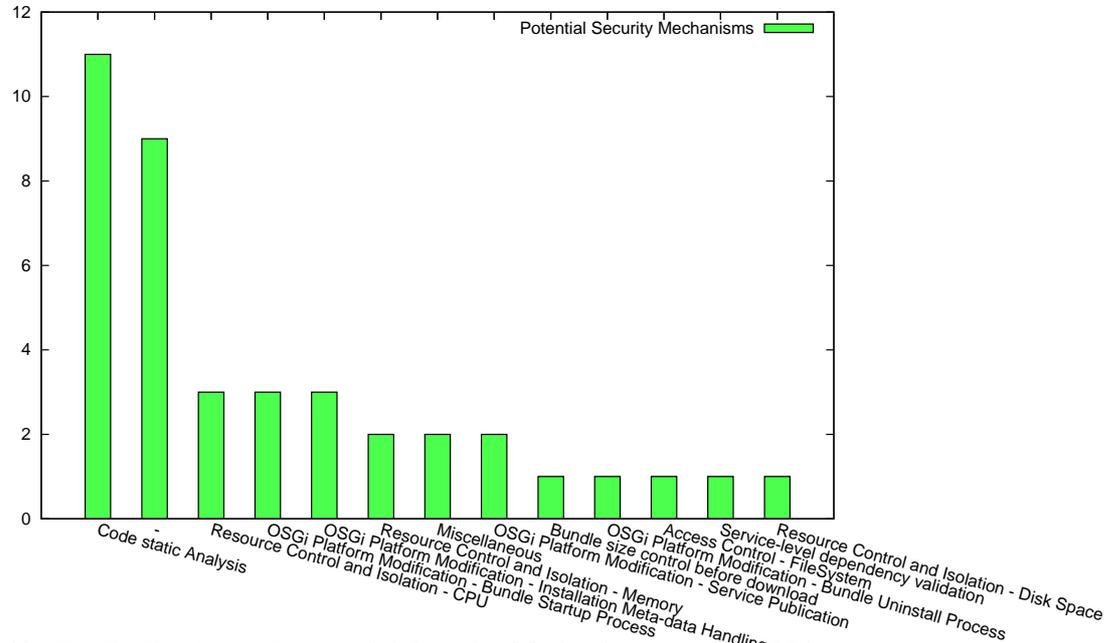, width=330pt}
\caption{Cardinality for each potential Security Mechanisms for the OSGi platform}
\label{fig:protections-card}
\end{figure}

The Figure \ref{fig:protections-card} shows the potential protections identified to protect an OSGi Platform against the
considered attacks. This potential protections are off course only set as hypotheses: as long as no implementation is available, it is not possible to assert that no special case or hard-to-track false positives and negatives dot not occur if the propose technique is used. The most promising approach seems to be static code analysis, that would help track both dangerous calls without heavyweight permissions and unsafe algorithms. The OSGi Platform itself would take benefit of several minor modifications: better handling of ill-formed meta-data, safe startup process for bundles, better control of service publication. Some of these mechanisms have been experienced in the SFelix Platform, and prove to be easy to implement. Also, resource control and isolation mechanisms (CPU, Memory, disk space) would make the support of multi-processes safer.

The potential protection mechanism represent the element that are worth an important development effort. However, they do not show the relative priority of the security mechanisms. Urgent security mechanisms are the ones that prevent attacks with serious consequences - for instance platform unavailability - to be performed, or the ones that are required to make the use of existing mechanisms efficient and convenient for developers. Consequently, the priority is to be set on following protection mechanisms:

\begin{itemize*}
 \item Protection of Attacks targeted at the whole Platform (see Figure \ref{fig:targets}), and that impair the availability or the performance of all executed bundles simultaneously,
 \item Protection against silent attacks: classical access control mechanisms are required inside the OSGi Platform, to support mutually untrustful bundles,
 \item tools are required to take advantages of existing mechanisms: for instance, Permission are supported, but currently extremely unconvenient to set and manage.
\end{itemize*}

We presented the requirement for developing a Hardened OSGi Platform by identifying the best promising potential security mechanisms as well as the most urgent tools for preventing serious attacks, or taking advantage of existing protections. However, developers require ready-to-use guidelines to take advantage of the knowledge we gathered in these study: in the absence of available tools, they have to take care by themselves that the code they produce is safe from the known vulnerabilities.

\subsection{Recommendations for a Hardened Execution Environment}
\label{subsec:specifications}

\paragraph{Hardening the Specifications of the OSGi Platform}
Based on the identified vulnerability of the OSGi Platform, we propose following recommendation for an enhanced OSGi Platform. These recommendation do not pretend to solve every identified problems, but intend to make the community aware of the easy changes that can be made to the OSGi Specification so as to prevent avoidable flaws.

These recommendations are validated by the Platform SFelix version 0.2, which is a robust extension to the Felix 0.8.0 implementation of the OSGi Platform.

Following improvement to the OSGi Release 4 Specification should be made:

\begin{itemize*}
  \item \textbf{Bundle Installation Process:} a maximum storage size for bundle archives is set. Alternatively, a maximum storage size for all data stored on the local disk is set (\emph{Bundle Archives and files created by the bundles}); \emph{OSGi R4 par. 4.3.3}.
  \item \textbf{Bundle Uninstallation Process:} remove the data on the local bundle filesystem when a bundle is uninstalled (and not when the platform is stopped); \emph{OSGi R4 par. 4.3.8}.
  \item \textbf{Bundle Signature Validation Process:} the digital signature must be checked at installed time. It must not rely on the Java built-in validation mechanism, since this latter is not compliant with the OSGi R4 Specifications \cite{parrend2007sfelix}; \emph{OSGi R4, Paragraph 2.3}.
  \item \textbf{Bundle Dependency Resolution Process:} do not reject duplicate imports. just ignore them; \emph{OSGi R4 par. 3.5.4}.
  \item \textbf{Bundle Start Process:} launch the Bundle Activator in a separate thread; \emph{OSGi R4 par. 4.3.5}.
  \item \textbf{OSGi Service Registration:} set a Platform Property that explicitly limits the number of registered services (default could be 50); \emph{OSGi R4 par. 5.2.3}.
  \item \textbf{Bundle Download:} when a bundle download facility is available, the total size of the bundles to installed should be checked immediately after the dependency resolution process. The bundles should be installed only if the required storage is available.
\end{itemize*}

To support this modifications of the OSGi R4 specifications, following changes have been applied to the API:

\begin{itemize}
  \item In the Class BundleContext, a method `getAvailableStorage()' is defined,
  \item A property `osgi.storage.max' is defined, that is set in the property configuration file of the OSGi framework.
  \item In the class org.osgi.service.obr.Resource, a method `getSize()' is defined. This method relies on the `size' entry of the bundle meta-data representation (usually  a XML file).
\end{itemize}

In addition to these simple enhancement, more research work is required in order to define proper solution to the identified vulnerabilities. The most important ones are the following:

\begin{itemize*}
  \item Static Code Analysis for Java,
  \item Convenient Permission Management for Java and OSGi,
  \item Resource isolation in component systems,
  \item Mandatory Service Management.
\end{itemize*}

Through this study, we identified both technical requirements for enhancement of the OSGi R4 Specifications, and necessary research work that is necessary to protect the OSGi Platform.

\paragraph{Hardening the Specifications of the Java Virtual Machine}
Some safety requirements have also been identified at the Virtual Machine Level.

The flaws that have been identified in the Sun Java Virtual Machine version 1.6 are the following:
\begin{itemize}
 \item the Java Permission `exitVM' appears not to be effective,
 \item the presence of a manifest with a huge size in a loaded Jar file introduces a dramatic slowdown of the JVM when the archive Manifest is extracted. Our implementation shows that a simple patch can correct this matter of fact.
\end{itemize}

A flaw has also been identified in the Gnu Classpath, which is an open source implementation of the Java classes. Gnu Classpath is used in conjunction with the JamVM Virtual Machine and targets resource-limited devices:
\begin{itemize}
 \item the presence of a manifest with a huge size in a loaded Jar file introduces a dramatic slowdown of the JVM when the corresponding JarFile Object is created, even though the Manifest stays unused.
\end{itemize}

\bigskip

Requirements for programming secure OSGi Systems have been identified. First, a hardened version of the OSGi Platform is needed to prevent most of the identified vulnerabilities to be exploited. However, since such a platform will take time to develop and validate, a pragmatic approach is to be taken. First, tools should be developed to ease the management of current security mechanisms such as Java Permissions, which are currently not adapted to dynamic systems. Secondly, developers need to keep on mind what the OSGi vulnerabilities are: this is made possible by the \emph{Seven Deadly Sins} of the OSGi R4 Platform.

%% file: tables/PlatformVulnerabilities.tex
\begin{table}
\begin{center}
\begin{tabular}{|p{150pt}|c|c|c|c|c|p{60pt}|}
\hline
\textbf{Vulnerability}&\textbf{Felix}&\textbf{Knopflerfish}&\textbf{Equinox}&\textbf{Concierge}&\textbf{SFelix}&\textbf{Any with Java Permissions}\\\hline
Exponential Object Creation&V&V&V&V&V&-\\\hline
Excessive Size of Manifest File&V&V&R&V&R&-\\\hline
Access Protected Package through split Packages&V&V&V&-&V&R\\\hline
Freezing Numerous Service Registration&-&-&-&V&-&-\\\hline
Big File Creator&V&V&V&V&V&R\\\hline
Management Utility Freezing - Thread Hanging&V&V&V&V&R&-\\\hline
Erroneous values of Manifest attributes&V&V&V&V&V&-\\\hline
Invalid Digital Signature Validation&V&-&-&-&R&-\\\hline
Cycle Between Services&V&V&V&V&V&-\\\hline
Hanging Thread&V&V&V&V&V&-\\\hline
Management Utility Freezing - Infinite Loop&V&V&V&V&R&-\\\hline
Fragment Substitution&V&V&V&-&V&R\\\hline
Numerous Service Registration&V&V&V&V&R&-\\\hline
Sleeping Bundle&V&V&V&V&V&-\\\hline
Code Observer&V&V&V&-&V&R\\\hline
Recursive Thread Creation&V&V&V&V&V&-\\\hline
Duplicate Package Import&V&V&R&R&R&-\\\hline
Memory Load Injection&V&V&V&V&V&-\\\hline
Infinite Loop in Method Call&V&V&V&V&V&-\\\hline
Runtime.halt&V&V&V&V&V&R\\\hline
CPU Load Injection&V&V&V&V&V&R\\\hline
System.exit&V&V&V&V&V&R\\\hline
Runtime.exec.kill&V&V&V&V&V&R\\\hline
Component Data Modifier&V&V&V&V&V&R\\\hline
Stand Alone Infinite Loop&V&V&V&V&V&-\\\hline
Execute Hidden Classes&V&V&V&-&V&R\\\hline
Pirat Bundle Manager&V&V&V&V&V&R\\\hline
Big Component Installer&-&-&-&-&R&-\\\hline
Launch a Hidden Bundle&V&V&V&V&V&R\\\hline
Zombie Data&V&R&R&V&R&-\\\hline
Decompression Bomb&-&-&-&-&-&-\\\hline
Hidden Method Launcher&V&V&V&V&V&R\\\hline
\multicolumn{7}{|l|}{V: Platform is Vulnerable; R: Platform is Robust; - : not relevant} \\\hline
\end{tabular}
\end{center}
\caption{Vulnerabilities for the main Open Source OSGi Platforms}
\label{tab:platform_vulnerabilities}
\end{table}

%% file: main/conclusions.tex
\section{Conclusions}
\label{sec:conclusions}

The objective of our study is to improve the dependability level of the OSGi platform, as well as the knowledge that is available relative to the vulnerabilities of the OSGi Platform. This improvement is achieved through four complementary contributions. First, we define a method for analyzing the security status of software systems, based on a specific Software Vulnerability Pattern. Secondly, we provide a vulnerability catalog that identified a set of vulnerabilities, and the key properties for understanding - and preventing - them. Thirdly, we developed a hardened OSGi Platform, SFelix v0.2\footnote{http://sfelix.gforge.inria.fr/}, that provides proof of concept protection mechanisms. And we issue a set of recommendations for the OSGi Specifications that integrate these protection mechanisms.

Our study is centered on the OSGi Core specification, and does not take into account several mechanisms that are - or can be - often used together with OSGi platforms. In particular, management facilities, such as JVMTI, and with less impact JMX have not been studied. OSGi standard services are neither been considered, and service engineering questions have been neglected. These three elements will require further work, and will likely enrich our vulnerability catalog.

A side-effect achievement of our study is to precisely identify the requirements in term of research and development, so as to provide OSGi platform that are actually robust, and not just partially hardened. Static Code analysis seem to be very promising, but suffers from significant theoretical limitation, especially in the world of Object-Oriented Languages. Convenient permission management, and proper resource isolation in Java multi-application systems are also a strong need on the road toward OSGi security. 

The present study provides a pragmatic approach to software security concerns, targeted at the world of OSGi Platforms. It present an important step toward a better understanding of OSGi-related security, and help practitioners implement safer system by providing a hardened OSGi prototype, SFelix v0.2. An important research effort is still required to provide an OSGi platform which security mechanisms can be said to be complete.

%% file: appendix/osgiPlatform.tex
\section{The OSGi platform}
\label{sec:osgiPlatform}

The OSGi Platform\footnote{http://www.osgi.org/} \cite{osgi05core} is a componentization layer to the Java Virtual Machine. It supports the runtime extension of Java-based application through a modular approach: the applications are parted into `bundles', that can be loaded, installed and managed independently from each other. 

In this section, we present first an overview of the OSGi Platform, then the core concept of OSGi: the bundles and their Life Cycle, and the possible interactions between bundles.

\subsection{Overview}

The OSGi Platform has been developed so as to support extensible Java-based systems in resource-constraint systems, such as automotive and mobile environments. It has since then spread into the world of Integrated Development Applications (in particular with Eclipse), and into applicative servers (IBM Websphere 6.1, Harmony, Cocoon, Directory...). 

\begin{figure}[htb]
\centering
\epsfig{file=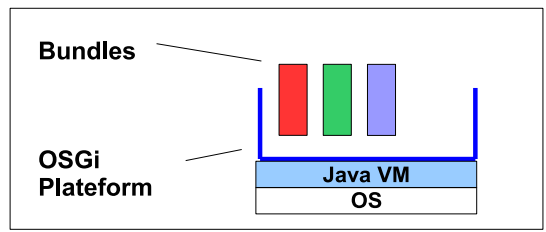, width=200pt}
\caption{Overview of an OSGi Platform}
\label{fig:osgi-overview}
\end{figure}

It runs as an overlay to the Java Virtual Machine (JVM). The figure \ref{fig:osgi-overview} shows the overview of an OSGi-based system, with the Operating System (OS), the JVM, the platform itself, and the bundles it contains.

Three main concepts sustain the OSGi platform: the platform, the bundle, and the interoperability between the bundles. The Platform manages the applications. The bundles are the unit of deployment and execution. The interoperability between the bundles is achieved at the class level (access to packages from other bundles) and at the service level (access to services registered by other bundles).

\subsection{The Bundles}

\begin{figure}[htb]
\centering
\epsfig{file=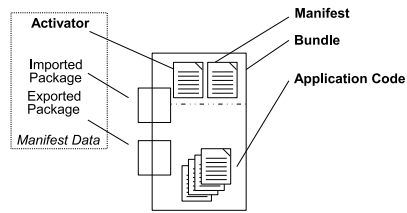, width=150pt}
\caption{Intern Structure of an OSGi bundle}
\label{fig:osgi-bundle}
\end{figure}

An OSGi bundle is a Jar file \cite{jarSpec} which is enhanced by specific meta-data. The typical structure of a bundle is shown in the figure \ref{fig:osgi-bundle}. The META-INF/MANIFEST.MF file contains the necessary OSGi meta-data: the bundle reference name (the `symbolic name'), its version, the dependencies and the provided resources. Some packages are exported, \emph{i.e.} accessible from other bundles inside the platform. The activator is used by the platform as an initial bootstrap when the bundle is started. Packages can be exported. Services can be registered, so as to be available for other bundles.

Each bundle has a restricted view on the OSGi platform: the OSGi Context, which is transmitted to the bundle activator at start time.
This context reference is needed to publish and look-up for services. It also supports the access to the management functionalities of the platform.

The OSGi bundles can also access the Operating System of the machine it is running on through native libraries. This possibility is not specific to the OSGi environment, since it relies on the Java Runtime API, but it allows the bundles to break their isolation.

\begin{figure}[htb]
\centering
\epsfig{file=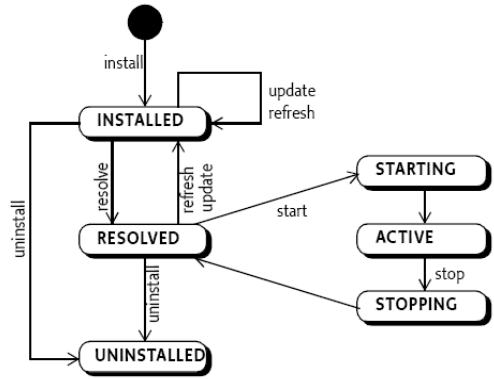, width=200pt}
\caption{Life Cycle of an OSGi Bundles inside the platform}
\label{fig:bundle-lf}
\end{figure}

The Life Cycle of a bundle inside the OSGi Platform is defined as follows. The bundle must first be installed. When it is required to start, the package-level dependencies with other bundles are resolved. When all dependencies are resolved, the bundle activator is launched: the \textsf{start()} method is called, and the related code is executed. Typically, these operations consist in configuration and publication of services.
 The bundle is then in the `started' state. Updating, stopping and uninstalling build the last possible operations for bundle management The figure \ref{fig:bundle-lf} shows the Life Cycle of a bundle inside a OSGi Platform.

\subsection{Interactions between Bundles}

\begin{figure}[htb]
\centering
\epsfig{file=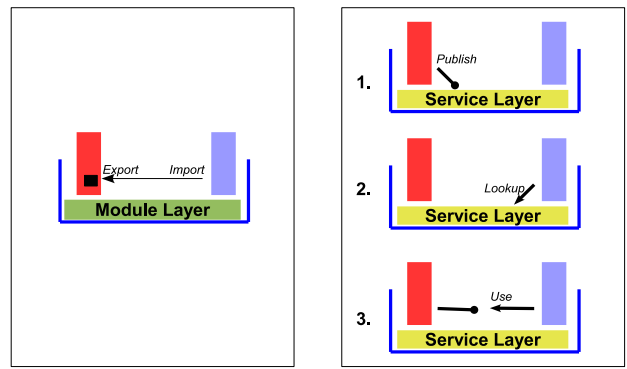, width=230pt}
\caption{Interaction Mechanisms between the OSGi Bundles}
\label{fig:osgi-interoperability}
\end{figure}

The interactions between the bundles are done through two complementary mechanisms: the package export/import and the service registration lookup facility. These mechanisms are shown in the figure \ref{fig:osgi-interoperability}.

The publication and lookup of services are performed through the BundleContext reference that each bundle receives ar startup time. During the publication process, the advertising bundles registers a service by publishing a Java interface it is implementing, and by providing a class implementing this interface. The lookup is performed by the client bundle, which gets the service from the BundleContext and uses it as a standard Java object.

%% file: appendix/vulnerabilityLists.tex
\section{Vulnerabilities List}

The most common Vulnerability Lists presented in the section \ref{subsec:classification} are given here.

\subsection{The Lindqvist Classification}
\label{sec:lindqvist}

The computer security intrusions identified by Lindqvist \cite{Lindqvist1997classification} are the following:

\begin{itemize*}
 \item external misuse (not technical), 
 \item hardware misuse, 
 \item masquerading, 
 \item setting up subsequent misuse, 
 \item bypassing intended controls, 
 \item active misuse of resource, 
 \item passive misuse of resource, 
 \item misuse resulting from inaction, 
 \item use of an indirect aid in committing other misuse.
\end{itemize*}

\subsection{Common Weaknesses Enumeration (CWE)}
\label{sec:cwe}

The categories defined in the Common Weaknesses Enumeration \cite{martin2005enumeration} are the following:

\begin{itemize*}
 \item Buffer overflows, format strings, etc. [BUFF];
 \item Structure and Validity Problems;[SVM];
 \item Special Elements [SPEC]; 
 \item Common Special Element Manipulations[SPECM]; 
 \item Technology-Specific Special Elements[SPECTS]; 
 \item Pathname Traversal and Equivalence Errors [PATH]; 
 \item Channel and Path Errors [CP]; 
 \item Information Management Errors [INFO]; 
 \item Race Conditions [RACE]; 
 \item Permissions, Privileges, and ACLs [PPA];
 \item Handler Errors [HAND];
 \item User Interface Errors [UI];
 \item Interaction Errors [INT];
 \item Initialization and Cleanup Errors [INIT]; 
 \item Resource Management Errors [RES];
 \item Numeric Errors [NUM];
 \item Authentication Error [AUTHENT];
 \item Cryptographic errors [CRYPTO];
 \item Randomness and Predictability [RAND];
 \item Code Evaluation and Injection [CODE];
 \item Error Conditions, Return Values, Status Codes [ERS];
 \item Insufficient Verification of Data [VER];
 \item Modification of Assumed-Immutable Data [MAID];
 \item Product-Embedded Malicious Code [MAL];
 \item Common Attack Mitigation Failures [ATTMIT];
 \item Containment errors (container errors) [CONT];
 \item Miscellaneous WIFFs [MISC].
\end{itemize*}

\subsection{Nineteen Dealy Sins}
\label{sec:19}

The 19 Deadly Sins defined by Howard \cite{Howard2005_19sins} are the following:

\begin{itemize*}
 \item buffer overflows,
 \item command injection,
 \item Cross-site scripting (XSS),
 \item format string problems, 
 \item integer range error, 
 \item SQL injection, 
 \item trusting network address information, 
 \item failing to protect network traffic, 
 \item failing to store and protect data, 
 \item failing to use cryptographically strong random numbers, 
 \item improper file access, 
 \item improper use of SSL, 
 \item use of weak password-based systems, 
 \item unauthenticated key exchange, 
 \item signal race condition, 
 \item use of 'magic' URLs and hidden forms, 
 \item failure to handle errors, 
 \item poor usability, 
 \item information leakage.
\end{itemize*}

\subsection{OWASP Top Ten}
\label{sec:owasp}

The OWASP Top Ten Vulnerability list for 2007 is the following\footnote{http://www.owasp.org/index.php/Top\_10\_2007-WIKI-FORMAT-TEST}: 

\begin{itemize*}
 \item Cross Site Scripting (XSS)
 \item Injection Flaws
 \item Malicious File Execution
 \item Insecure Direct Object Reference
 \item Cross Site Request Forgery (CSRF)
 \item Information Leakage and Improper Error Handling
 \item Broken Authentication and Session Management
 \item Insecure Cryptographic Storage
 \item Insecure Communications
 \item Failure to Restrict URL Access
\end{itemize*}

\subsection{Seven Kingdoms}
\label{sec:7}

The Seven Kingdoms defined by Gary McGraw \cite{mcgraw06softwaresecurity} are the following. Note that each Kingdom contains a certain number of Phyla, that help give more precise hints so as the actual vulnerabilities.

\begin{itemize*}
 \item Input Validation and representation, 
 \item API abuse, 
 \item Security Features, 
 \item Time and state, 
 \item error handling, 
 \item code quality, 
 \item encapsulation 
 \item + environment
\end{itemize*}

%% file: appendix/formalVulnerabilityPattern.tex
\section{Formal Expression of the Vulnerability Pattern}
\label{app:formalVulnerabilityPattern}

This section presents the Vulnerability Pattern in the Augmented Backus Naur Form (BNF) \cite{rfc4234}. 

The current grammar is not meant to be closed: it reflects the knowledge relative to the considered vulnerabilities at a given time. It can be extended with additional attribute values.

The catalog of the OSGi Malicious Bundles is referred as the `mb' catalog. 

\paragraph{Vulnerability Reference}

\begin{itemize*}
  \item VULNERABILITY\_NAME ::= text
  \item IDENTIFIER ::= CATALOG\_ID.SRC\_REF.ID\\
with:\\
CATALOG\_ID ::= mb \\
SRC\_REF ::= archive|java|native|osgi\\
ID ::= (0-9)*
  \item ORIGIN ::= text
  \item LOCATION ::= Bundle ( Archive | Manifest | Activator | Fragment ) | Application Code - ( Native Code | Java ( Code | API ) | OSGi API )
  \item SOURCE ::= (ENTITY { ( FUNCTIONNALITY | FLAW  ;)+};)+\\
with ENTITY ::= OS | JVM - ( Runtime API | APIs )| OSGi Platform - (( Module | Life-Cycle | Service ) Layer | Bundle Repository Client )| Application Code\\
FUNCTIONNALITY ::= Kill utility | Value of Method Parameters | ( System.exit | Runtime.halt ) method | Native Code Execution | Thread API | Reflection API | ClassLoader API | File API | Java Archive | Bundle Management | Bundle Fragments
and:\\
FLAW ::= No Algorithm Safety - ( Java | Native Code )| Non OSGi R4-compliant Digital Signature Validation in the JVM | No Verification of Bundle Archive Validity | No Check of Size of Loaded Bundles | No Check of Size of stored Data | No safe Bundle Start | No Removal of Uninstalled Bundle Data | Bundle Meta-data Handling - No Safe-Default | Uncontrolled Service Registration | Architecture of the Application - No Validation of Service Dependency
  \item TARGET ::= Platform | OSGi Element - ( Platform Management Utility | Bundle | Service|Package )
  \item CONSEQUENCE\_TYPE ::= ( Unavailability | Performance Breakdown | Undue Access )( - ( Platform | Service | Package )(, ( Platform | Service | Package ))*)?
  \item INTRODUCTION\_TIME ::= Platform Design or Implementation | Development | Bundle Meta-data Generation | Bundle Digital Signature | Installation | Service Publication or Resolution
  \item EXPLOIT\_TIME ::=  Download | Installation | Bundle Start | Execution 
\end{itemize*}

\paragraph{Vulnerability Description}

\begin{itemize*}
  \item DESCRIPTION ::= text
  \item PRECONDITIONS ::= text 
  \item ATTACK\_PROCESS ::= text
  \item CONSEQUENCE\_DESCRIPTION ::= text
  \item SEE\_ALSO ::= VULNERABILITY\_NAME (, VULNERABILITY\_NAME)*
\end{itemize*}

\paragraph{Vulnerability Implementation}

\begin{itemize*}
  \item CODE\_REFERENCE ::= FILE\_NAME\\
with FILE\_NAME the name of a file, as defined by Unix File Names
  \item OSGI\_PROFILE ::= CDC-1.0/Foundation-1.0 | OSGi/Minimum-1.1 | JRE-1.1 | J2SE-1.2 | J2SE-1.3 | J2SE-1.4 | J2SE-1.5 | J2SE-1.6 | PersonalJava-1.1 | PersonalJava-1.2 | CDC-1.0/PersonalBasis-1.0 | CDC-1.0/PersonalJava-1.0
  \item DATE ::= MONTH.DAY.YEAR\\
with MONTH ::= (1-12), DAY ::= (1-31), YEAR ::= (0-3000)
  \item TEST\_COVERAGE ::=  (0-100)  \%
  \item TESTED\_ON ::= Oscar | Felix | Knopflerfish | Equinox
\end{itemize*}

\paragraph{Protection}

\begin{itemize*}
  \item EXISTING\_MECHANISMS ::= Java Permissions | OSGi AdminPermission | SFelix OSGi Security Layer | - 
  \item ENFORCEMENT\_POINT ::= Platform startup | Bundle Installation | - 
  \item POTENTIAL\_MECHANISMS ::= (POTENTIAL\_MECHANISM\_NAME (POTENTIAL\_MECHANISM\_DESCR)?)+
with POTENTIAL\_MECHANISM\_NAME ::= Code static Analysis | OSGi Platform Modification - ( Bundle Startup Process | Installation Meta-data Handling | Service Publication )| Bundle size control before download | Service-level dependency validation | Resource Control and Isolation - ( CPU | Memory | Disk Space )| Access Control - FileSystem | Miscellaneous | -
and POTENTIAL\_MECHANISM\_DESCR ::= text
  \item ATTACK\_PREVENTION ::= Stop a ill-behaving thread | -
  \item REACTION ::= Uninstall the malicious bundle | Erase files | Stop the system process | Restart the platform | -
\end{itemize*}

%% file: catalog.tex
\section{Vulnerability Catalog}
\label{sec:catalog}

\subsection{Bundle Archive}

\input{vulnerabilityPatterns/invalidDigitalSignatureValidation}

\newpage
\input{vulnerabilityPatterns/bigComponentInstaller}

\newpage
\input{vulnerabilityPatterns/decompressionBomb}

\newpage
\subsection{Bundle Manifest}

\input{vulnerabilityPatterns/duplicatePackageImport}

\newpage
\input{vulnerabilityPatterns/excessiveManifestSize}

\newpage
\input{vulnerabilityPatterns/erroneousManifestEntryValues}

\newpage
\subsection{Bundle Activator}

\input{vulnerabilityPatterns/managementUtilityFreezingInfiniteLoop}

\label{sec:managementFreezingLoop}

\newpage
\input{vulnerabilityPatterns/managementUtilityFreezingHanging}

\newpage
\subsection{Bundle Code - Native}

\input{vulnerabilityPatterns/exec.kill}

\newpage
\input{vulnerabilityPatterns/cpuLoadInjection}

\newpage
\subsection{Bundle Code - Java}

\input{vulnerabilityPatterns/system.exit}

\newpage
\input{vulnerabilityPatterns/runtime.halt}

\newpage
\input{vulnerabilityPatterns/recursiveThreadCreation}

\newpage
\input{vulnerabilityPatterns/hangingThread}

\newpage
\input{vulnerabilityPatterns/sleepingBundle}

\newpage
\input{vulnerabilityPatterns/bigFileCreator}

\newpage
\input{vulnerabilityPatterns/codeObserver}

\newpage
\input{vulnerabilityPatterns/componentDataModifier}

\newpage
\input{vulnerabilityPatterns/hiddenMethodLauncher}

\newpage
\input{vulnerabilityPatterns/memoryLoadInjection}

\newpage
\input{vulnerabilityPatterns/standAloneInfiniteLoop}

\newpage
\input{vulnerabilityPatterns/infiniteLoopInMethodCall}

\newpage
\input{vulnerabilityPatterns/exponentialObjectCreation}

\newpage
\subsection{Bundle Code - OSGi APi}

\input{vulnerabilityPatterns/launchHiddenBundle}

\newpage
\input{vulnerabilityPatterns/piratBundleManager}

\newpage
\input{vulnerabilityPatterns/zombieData}

\newpage
\input{vulnerabilityPatterns/cycleBetweenServices}

\newpage
\input{vulnerabilityPatterns/numerousServiceRegistration}

\newpage
\input{vulnerabilityPatterns/freezingNumerousServiceRegistration}

\newpage
\subsection{Bundle Fragments}

\input{vulnerabilityPatterns/executeHiddenClasses}

\newpage
\input{vulnerabilityPatterns/fragmentSubstitution}

\newpage
\input{vulnerabilityPatterns/splitPackage}

%% file: vulnerabilityPatterns/invalidDigitalSignatureValidation.tex
\subsubsection{Invalid Digital Signature Validation}

\paragraph{Vulnerability Reference}

\begin{itemize*}
  \item \textbf{Vulnerability Name:} Invalid Digital Signature Validation
  \item \textbf{Identifier:} Mb.archive.1
  \item \textbf{Origin:} Ares research project `malicious-bundle'
  \item \textbf{Location of Exploit Code:} Bundle Archive
  \item \textbf{Source:} OSGi Platform - Life-Cycle Layer (Non OSGi R4-compliant Digital Signature Validation in the JVM)
  \item \textbf{Target:} Platform
  \item \textbf{Consequence Type:} Undue Access
  \item \textbf{Introduction Time:} Bundle Digital Signature
  \item \textbf{Exploit Time:} Installation
\end{itemize*}

\paragraph{Vulnerability Description}

\begin{itemize*}
  \item \textbf{Description:} A bundle which signature is NOT compliant to the OSGi R4 Digital Signature is installed on the platform
  \item \textbf{Preconditions:} No Digital Signature Validation, or Digital Signature Validation Process that relies on the Java JarFile API to perform the validation of the digital signature. The bundle signature must be non OSGi R4 compliant in one of the following ways: resources have been removed from the archive; resources have been added; the first resources in the archive are NOT the Manifest File, the Signature File and the Signature Block file in this order (see \cite{parrend06deployment}).
  \item \textbf{Attack Process:} Install a bundle with an invalid digital signature (see preconditions)
  \item \textbf{Consequence Description:} -
  \item \textbf{See Also:} -
\end{itemize*}

\paragraph{Protection}

\begin{itemize*}
  \item \textbf{Existing Mechanisms:} SFelix OSGi Security Layer
  \item \textbf{Enforcement Point:} Bundle Installation
  \item \textbf{Potential Mechanisms:} - 
  \item \textbf{Attack Prevention:} -
  \item \textbf{Reaction:} Uninstall the malicious bundle
\end{itemize*}

\paragraph{Vulnerability Implementation}

\begin{itemize*}
  \item \textbf{Code Reference:} Bindex-resourceRemoved-1.0.jar, bindex-resourcesAdded-1.0.jar, bindex-unvalidResourceOrder-1.0.jar
  \item \textbf{OSGi Profile:} J2SE-1.5
  \item \textbf{Date:} 2007-04-25
  \item \textbf{Test Coverage:} 100\%
  \item \textbf{Known Vulnerable Platforms:} Felix
  \item \textbf{Known Robust Platforms:} SFelix
\end{itemize*}

%% file: vulnerabilityPatterns/bigComponentInstaller.tex
\subsubsection{Big Component Installer}

\paragraph{Vulnerability Reference}

\begin{itemize*}
  \item \textbf{Vulnerability Name:} Big Component Installer
  \item \textbf{Identifier:} Mb.archive.2
  \item \textbf{Origin:} Ares research project `malicious-bundle'
  \item \textbf{Location of Exploit Code:} Bundle Archive
  \item \textbf{Source:} OSGi Platform - Bundle Repository Client (No Check of Size of Loaded Bundles)
  \item \textbf{Target:} Platform
  \item \textbf{Consequence Type:} Performance Breakdown
  \item \textbf{Introduction Time:} Platform Design or Implementation
  \item \textbf{Exploit Time:} Execution
\end{itemize*}

\paragraph{Vulnerability Description}

\begin{itemize*}
  \item \textbf{Description:} Remote installation of a bundle which size is of similar to the available device memory
  \item \textbf{Preconditions:} OSGi platform running on a memory limited device
  \item \textbf{Attack Process:} -
  \item \textbf{Consequence Description:} Little memory is available for subsequent operations
  \item \textbf{See Also:} Big File Creator
\end{itemize*}

\paragraph{Protection}

\begin{itemize*}
  \item \textbf{Existing Mechanisms:} OSGi AdminPermission
  \item \textbf{Enforcement Point:} -
  \item \textbf{Potential Mechanisms:} Bundle size control before download 
  \item \textbf{Attack Prevention:} -
  \item \textbf{Reaction:} -
\end{itemize*}

\paragraph{Vulnerability Implementation}

\begin{itemize*}
  \item \textbf{Code Reference:} -
  \item \textbf{OSGi Profile:} J2SE-1.5
  \item \textbf{Date:} 2007-02-20
  \item \textbf{Test Coverage:} 00\%
\end{itemize*}

%% file: vulnerabilityPatterns/decompressionBomb.tex
\subsubsection{Decompression Bomb}

\paragraph{Vulnerability Reference}

\begin{itemize*}
  \item \textbf{Vulnerability Name:} Decompression Bomb
  \item \textbf{Identifier:} Mb.archive.3
  \item \textbf{Origin:} Ares research project `malicious-bundle'
  \item \textbf{Location of Exploit Code:} Bundle Archive
  \item \textbf{Source:} OSGi Platform - Life-Cycle Layer (No Verification of Bundle Archive Validity)
  \item \textbf{Target:} Platform
  \item \textbf{Consequence Type:} Performance Breakdown
  \item \textbf{Introduction Time:} Development
  \item \textbf{Exploit Time:} Execution
\end{itemize*}

\paragraph{Vulnerability Description}

\begin{itemize*}
  \item \textbf{Description:} The Bundle Archive is a decompression Bomb (either a huge file made of identical bytes, or a recursive archive)
  \item \textbf{Preconditions:} -
  \item \textbf{Attack Process:} Provide a Bundle Archive that is a decompression Bomb for installation (on a OBR, etc.)
  \item \textbf{Consequence Description:} Important consumption of CPU or memory.
  \item \textbf{See Also:} -
\end{itemize*}

\paragraph{Protection}

\begin{itemize*}
  \item \textbf{Existing Mechanisms:} -
  \item \textbf{Enforcement Point:} -
  \item \textbf{Potential Mechanisms:} OSGi Platform Modification - Bundle Startup Process (Check that the Bundle is not a Decompression Bomb archive)
  \item \textbf{Attack Prevention:} -
  \item \textbf{Reaction:} -
\end{itemize*}

\paragraph{Vulnerability Implementation}

\begin{itemize*}
  \item \textbf{Code Reference:} Fr.inria.ares.decompressionbomb-0.1.jar
  \item \textbf{OSGi Profile:} J2SE-1.5
  \item \textbf{Date:} 2007-04-20
  \item \textbf{Test Coverage:} 50\%
\end{itemize*}

%% file: vulnerabilityPatterns/duplicatePackageImport.tex
\subsubsection{Duplicate Package Import}

\paragraph{Vulnerability Reference}

\begin{itemize*}
  \item \textbf{Vulnerability Name:} Duplicate Package Import
  \item \textbf{Identifier:} Mb.osgi.1
  \item \textbf{Origin:} Ares research project `malicious-bundle'
  \item \textbf{Location of Exploit Code:} Bundle Manifest
  \item \textbf{Source:} OSGi Platform - Module Layer (Bundle Meta-data Handling - No Safe-Default)
  \item \textbf{Target:} OSGi Element - Bundle
  \item \textbf{Consequence Type:} Unavailability
  \item \textbf{Introduction Time:} Bundle Meta-data Generation
  \item \textbf{Exploit Time:} Installation
\end{itemize*}

\paragraph{Vulnerability Description}

\begin{itemize*}
  \item \textbf{Description:} A package is imported twice (or more) according to manifest attribute 'Import-Package'. In the Felix and Knopflerfish OSGi implementations, the bundle can not be installed
  \item \textbf{Preconditions:} -
  \item \textbf{Attack Process:} -
  \item \textbf{Consequence Description:} -
  \item \textbf{See Also:} Excessive Size of Manifest File, Unvalid Activator Meta-data, Erroneous values of Manifest attributes, Insufficient User Meta-data
\end{itemize*}

\paragraph{Protection}

\begin{itemize*}
  \item \textbf{Existing Mechanisms:} -
  \item \textbf{Enforcement Point:} -
  \item \textbf{Potential Mechanisms:} OSGi Platform Modification - Installation Meta-data Handling (ignore the repeated imports during OSGi metadata analysis)
  \item \textbf{Attack Prevention:} -
  \item \textbf{Reaction:} -
\end{itemize*}

\paragraph{Vulnerability Implementation}

\begin{itemize*}
  \item \textbf{Code Reference:} Fr.inria.ares.duplicateimport-0.1.ja
  \item \textbf{OSGi Profile:} J2SE-1.5
  \item \textbf{Date:} 2006-10-28
  \item \textbf{Test Coverage:} 100\%
  \item \textbf{Known Vulnerable Platforms:} Felix;
	Knopflerfish
  \item \textbf{Known Robust Platforms:} Equinox;
        Concierge;
        SFelix
\end{itemize*}

%% file: vulnerabilityPatterns/excessiveManifestSize.tex
\subsubsection{Excessive Size of Manifest File}

\paragraph{Vulnerability Reference}

\begin{itemize*}
  \item \textbf{Vulnerability Name:} Excessive Size of Manifest File
  \item \textbf{Identifier:} Mb.osgi.2
  \item \textbf{Origin:} Ares research project `malicious-bundle'
  \item \textbf{Location of Exploit Code:} Bundle Manifest
  \item \textbf{Source:} OSGi Platform - Module Layer (Bundle Meta-data Handling - No Safe-Default)
  \item \textbf{Target:} OSGi Element - Bundle
  \item \textbf{Consequence Type:} Unavailability
  \item \textbf{Introduction Time:} Bundle Meta-data Generation
  \item \textbf{Exploit Time:} Installation
\end{itemize*}

\paragraph{Vulnerability Description}

\begin{itemize*}
  \item \textbf{Description:} A bundle with a huge number of (similar) package imports (more than 1 Mbyte)
  \item \textbf{Preconditions:} -
  \item \textbf{Attack Process:} Insert a big number of imports in the manifest file of the bundle
  \item \textbf{Consequence Description:} In the Felix and Knopflerfish implementations, the launcher process takes a long time (several minutes) to parse the metadata file
  \item \textbf{See Also:} Duplicate Package Import, Unvalid Activator Meta-data, Erroneous values of Manifest attributes, Insufficient User Meta-data
\end{itemize*}

\paragraph{Protection}

\begin{itemize*}
  \item \textbf{Existing Mechanisms:} -
  \item \textbf{Enforcement Point:} -
  \item \textbf{Potential Mechanisms:} OSGi Platform Modification - Installation Meta-data Handling (check the size of manifest before the installation; more generally, check the format of the manifest size before the installation)
  \item \textbf{Attack Prevention:} -
  \item \textbf{Reaction:} -
\end{itemize*}

\paragraph{Vulnerability Implementation}

\begin{itemize*}
  \item \textbf{Code Reference:} Fr.inria.ares.hugemanifest-0.1.jar
  \item \textbf{OSGi Profile:} J2SE-1.5
  \item \textbf{Date:} 2006-10-28
  \item \textbf{Test Coverage:} 100\%
  \item \textbf{Known Vulnerable Platforms:} Felix;
	Knopflerfish;
	Concierge
  \item \textbf{Known Robust Platforms:} SFelix;
        Equinox
\end{itemize*}

%% file: vulnerabilityPatterns/erroneousManifestEntryValues.tex
\subsubsection{Erroneous values of Manifest attributes}

\paragraph{Vulnerability Reference}

\begin{itemize*}
  \item \textbf{Vulnerability Name:} Erroneous values of Manifest attributes
  \item \textbf{Identifier:} Mb.osgi.3
  \item \textbf{Origin:} Ares research project `malicious-bundle'
  \item \textbf{Location of Exploit Code:} Bundle Manifest
  \item \textbf{Source:} OSGi Platform - Module Layer (Bundle Meta-data Handling - No Safe-Default)
  \item \textbf{Target:} OSGi Element - Bundle
  \item \textbf{Consequence Type:} Unavailability
  \item \textbf{Introduction Time:} Bundle Meta-data Generation
  \item \textbf{Exploit Time:} Installation
\end{itemize*}

\paragraph{Vulnerability Description}

\begin{itemize*}
  \item \textbf{Description:} A bundle that provides false meta-data, in this example an non existent bundle update location
  \item \textbf{Preconditions:} -
  \item \textbf{Attack Process:} Set a false value for a given meta-data entry
  \item \textbf{Consequence Description:} The actions that rely on the meta-data can not be executed (here, no update possible)
  \item \textbf{See Also:} Duplicate Import, Excessive Size of Manifest File
\end{itemize*}

\paragraph{Protection}

\begin{itemize*}
  \item \textbf{Existing Mechanisms:} -
  \item \textbf{Enforcement Point:} -
  \item \textbf{Potential Mechanisms:} OSGi Platform Modification - Installation Meta-data Handling (check the format of the manifest size before the installation, and provide failsafe default)
  \item \textbf{Attack Prevention:} -
  \item \textbf{Reaction:} -
\end{itemize*}

\paragraph{Vulnerability Implementation}

\begin{itemize*}
  \item \textbf{Code Reference:} Fr.inria.ares.malformedupdatelocation-0.1.jar
  \item \textbf{OSGi Profile:} J2SE-1.5
  \item \textbf{Date:} 2006-10-28
  \item \textbf{Test Coverage:} 10\%
  \item \textbf{Known Vulnerable Platforms:} Felix;
	Equinox;
	Knopflerfish;
	Concierge;
	SFelix
\end{itemize*}

%% file: vulnerabilityPatterns/managementUtilityFreezingInfiniteLoop.tex
\subsubsection{Management Utility Freezing - Infinite Loop}

\paragraph{Vulnerability Reference}

\begin{itemize*}
  \item \textbf{Vulnerability Name:} Management Utility Freezing - Infinite Loop
  \item \textbf{Extends:} Infinite Loop in Method Call
  \item \textbf{Identifier:} Mb.osgi.4
  \item \textbf{Origin:} Ares research project `malicious-bundle'
  \item \textbf{Location of Exploit Code:} Bundle Activator
  \item \textbf{Source:} OSGi Platform - Life-Cycle Layer (No safe Bundle Start)
  \item \textbf{Target:} OSGi Element - Platform Management Utility
  \item \textbf{Consequence Type:} Performance Breakdown;
	Unavailability
  \item \textbf{Introduction Time:} Development
  \item \textbf{Exploit Time:} Bundle Start
\end{itemize*}

\paragraph{Vulnerability Description}

\begin{itemize*}
  \item \textbf{Description:} An infinite loop is executed in the Bundle Activator
  \item \textbf{Preconditions:} -
  \item \textbf{Attack Process:} An infinite loop is executed in the Bundle Activator
  \item \textbf{Consequence Description:} Block the OSGi Management entity (the felix, equinox or knopflerfish shell; when launched in the KF graphical interface, the shell remain available but the GUI is frozen). Because of the infinite loop, most CPU resource is consumed
  \item \textbf{See Also:} CPU Load Injection, Infinite Loop in Method Call, Stand Alone Infinite Loop, Hanging Thread
\end{itemize*}

\paragraph{Protection}

\begin{itemize*}
  \item \textbf{Existing Mechanisms:} -
  \item \textbf{Enforcement Point:} -
  \item \textbf{Potential Mechanisms:} Code static Analysis ;
	Resource Control and Isolation - CPU ;
	OSGi Platform Modification - Bundle Startup Process (launch the bundle activator in a separate thread to prevent startup hanging)
  \item \textbf{Attack Prevention:} -
  \item \textbf{Reaction:} -
\end{itemize*}

\paragraph{Vulnerability Implementation}

\begin{itemize*}
  \item \textbf{Code Reference:} Fr.inria.ares.infiniteloopinmethodcall-0.1.jar
  \item \textbf{OSGi Profile:} J2SE-1.5
  \item \textbf{Date:} 2006-08-24
  \item \textbf{Test Coverage:} 10\%
  \item \textbf{Known Vulnerable Platforms:} Felix;
	Equinox;
	Knopflerfish;
	Concierge
  \item \textbf{Known Robust Platforms:} SFelix
\end{itemize*}

%% file: vulnerabilityPatterns/managementUtilityFreezingHanging.tex
\subsubsection{Management Utility Freezing - Thread Hanging}

\paragraph{Vulnerability Reference}

\begin{itemize*}
  \item \textbf{Vulnerability Name:} Management Utility Freezing - Thread Hanging
  \item \textbf{Extends:} Hanging Thread
  \item \textbf{Identifier:} Mb.osgi.5
  \item \textbf{Origin:} Ares research project `malicious-bundle'
  \item \textbf{Location of Exploit Code:} Bundle Activator
  \item \textbf{Source:} OSGi Platform - Life-Cycle Layer (No safe Bundle Start)
  \item \textbf{Target:} OSGi Element - Platform Management Utility
  \item \textbf{Consequence Type:} Unavailability
  \item \textbf{Introduction Time:} Development
  \item \textbf{Exploit Time:} Bundle Start
\end{itemize*}

\paragraph{Vulnerability Description}

\begin{itemize*}
  \item \textbf{Description:} A hanging thread in the Bundle Activator makes the management utility freeze
  \item \textbf{Preconditions:} -
  \item \textbf{Attack Process:} -
  \item \textbf{Consequence Description:} Block the OSGi Management entity (the felix, equinox or knopflerfish shell; when launched in the KF graphical interface, the shell remain available but the GUI is frozen).
  \item \textbf{See Also:} Management Utility Freezing - Infinite Loop, Hanging Thread
\end{itemize*}

\paragraph{Protection}

\begin{itemize*}
  \item \textbf{Existing Mechanisms:} -
  \item \textbf{Enforcement Point:} -
  \item \textbf{Potential Mechanisms:} OSGi Platform Modification - Bundle Startup Process (launch the bundle activator in a separate thread);
	Code static Analysis 
  \item \textbf{Attack Prevention:} -
  \item \textbf{Reaction:} -
\end{itemize*}

\paragraph{Vulnerability Implementation}

\begin{itemize*}
  \item \textbf{Code Reference:} Fr.inria.ares.hangingthread-0.1.jar, fr.inria.ares.hangingthread2-0.1.jar
  \item \textbf{OSGi Profile:} J2SE-1.5
  \item \textbf{Date:} 2006-08-24
  \item \textbf{Test Coverage:} 20\%
  \item \textbf{Known Vulnerable Platforms:} Felix;
	Equinox;
	Knopflerfish;
	Concierge
  \item \textbf{Known Robust Platforms:} SFelix
\end{itemize*}

%% file: vulnerabilityPatterns/exec.kill.tex
\subsubsection{Runtime.exec.kill}

\paragraph{Vulnerability Reference}

\begin{itemize*}
  \item \textbf{Vulnerability Name:} Runtime.exec.kill
  \item \textbf{Identifier:} Mb.native.1
  \item \textbf{Origin:} Ares research project `malicious-bundle'
  \item \textbf{Location of Exploit Code:} Application Code - Native Code
  \item \textbf{Source:} OS (Kill utility);
        JVM - Runtime API (Native Code Execution)
  \item \textbf{Target:} Platform
  \item \textbf{Consequence Type:} Unavailability
  \item \textbf{Introduction Time:} Development
  \item \textbf{Exploit Time:} Execution
\end{itemize*}

\paragraph{Vulnerability Description}

\begin{itemize*}
  \item \textbf{Description:} A bundle that stops the execution platform through an OS call
  \item \textbf{Preconditions:} No SecurityManager, or FilePermission `execute' on the required utilities (kill, ps, grep, cut)
  \item \textbf{Attack Process:} Kill the OS process which corresponds to the execution platform; this process is identified as far it is the parent process of the process in which the malicious script is executed
  \item \textbf{Consequence Description:} The shutdown hooks of the platforms arer executed
  \item \textbf{See Also:} System.exit, Runtime.halt, Recursive Thread Creation
\end{itemize*}

\paragraph{Protection}

\begin{itemize*}
  \item \textbf{Existing Mechanisms:} Java Permissions
  \item \textbf{Enforcement Point:} Platform startup
  \item \textbf{Potential Mechanisms:} Code static Analysis 
  \item \textbf{Attack Prevention:} -
  \item \textbf{Reaction:} Restart the platform
\end{itemize*}

\paragraph{Vulnerability Implementation}

\begin{itemize*}
  \item \textbf{Code Reference:} Fr.inria.ares.runtime\_exec\_kill-0.1.jar
  \item \textbf{OSGi Profile:} J2SE-1.5
  \item \textbf{Date:} 2006-08-21
  \item \textbf{Test Coverage:} 100\%
  \item \textbf{Known Vulnerable Platforms:} Felix;
	Equinox;
	Knopflerfish;
	Concierge;
	SFelix
\end{itemize*}

%% file: vulnerabilityPatterns/cpuLoadInjection.tex
\subsubsection{CPU Load Injection}

\paragraph{Vulnerability Reference}

\begin{itemize*}
  \item \textbf{Vulnerability Name:} CPU Load Injection
  \item \textbf{Identifier:} Mb.native.2
  \item \textbf{Origin:} MOSGI, Ares research project
  \item \textbf{Location of Exploit Code:} Application Code - Native Code
  \item \textbf{Source:} Application Code (No Algorithm Safety - Native Code);
        JVM - Runtime API (Native Code Execution)
  \item \textbf{Target:} Platform
  \item \textbf{Consequence Type:} Unavailability
  \item \textbf{Introduction Time:} Development
  \item \textbf{Exploit Time:} Execution
\end{itemize*}

\paragraph{Vulnerability Description}

\begin{itemize*}
  \item \textbf{Description:} A malicious bundle that consumes 80\% of the host CPU
  \item \textbf{Preconditions:} No SecurityManager, or RuntimePermission `loadLibrary'
  \item \textbf{Attack Process:} Execute a C call that consume CPU by switching between CPU-intensive calculation and sleep time, according to the specified ratio
  \item \textbf{Consequence Description:} Most of the available CPU of the system is consumed artificially
  \item \textbf{See Also:} Memory Load Injection, Ramping Memory Load Injection, Infinite Loop, Stand-alone Infinite Loop
\end{itemize*}

\paragraph{Protection}

\begin{itemize*}
  \item \textbf{Existing Mechanisms:} Java Permissions
  \item \textbf{Enforcement Point:} Platform startup
  \item \textbf{Potential Mechanisms:} Miscellaneous (extension of the Java-Level security mechanisms to the native code)
  \item \textbf{Attack Prevention:} -
  \item \textbf{Reaction:} Uninstall the malicious bundle
\end{itemize*}

\paragraph{Vulnerability Implementation}

\begin{itemize*}
  \item \textbf{Code Reference:} Fr.inria.ares.cpuloadinjector-0.1.jar
  \item \textbf{OSGi Profile:} J2SE-1.5
  \item \textbf{Date:} 2006-08-24
  \item \textbf{Test Coverage:} 00\%
  \item \textbf{Known Vulnerable Platforms:} Felix;
	Equinox;
	Knopflerfish;
	Concierge;
	SFelix
\end{itemize*}

%% file: vulnerabilityPatterns/system.exit.tex
\subsubsection{System.exit}

\paragraph{Vulnerability Reference}

\begin{itemize*}
  \item \textbf{Vulnerability Name:} System.exit
  \item \textbf{Identifier:} Mb.java.1
  \item \textbf{Origin:} Ares research project `malicious-bundle'
  \item \textbf{Location of Exploit Code:} Application Code - Java API
  \item \textbf{Source:} JVM - Runtime API (System.exit method)
  \item \textbf{Target:} Platform
  \item \textbf{Consequence Type:} Unavailability
  \item \textbf{Introduction Time:} Development
  \item \textbf{Exploit Time:} Execution
\end{itemize*}

\paragraph{Vulnerability Description}

\begin{itemize*}
  \item \textbf{Description:} A bundle that stops the platform by calling 'System.exit(0)'
  \item \textbf{Preconditions:} No SecurityManager, or presence of the RuntimePermission `exitVM'
  \item \textbf{Attack Process:} -
  \item \textbf{Consequence Description:} -
  \item \textbf{See Also:} Runtime.halt, Exec.Kill, Recursive Thread Creation
\end{itemize*}

\paragraph{Protection}

\begin{itemize*}
  \item \textbf{Existing Mechanisms:} Java Permissions
  \item \textbf{Enforcement Point:} Platform startup
  \item \textbf{Potential Mechanisms:} Code static Analysis 
  \item \textbf{Attack Prevention:} -
  \item \textbf{Reaction:} Restart the platform
\end{itemize*}

\paragraph{Vulnerability Implementation}

\begin{itemize*}
  \item \textbf{Code Reference:} Fr.inria.ares.system\_exit-0.1.jar
  \item \textbf{OSGi Profile:} J2SE-1.5
  \item \textbf{Date:} 2006-08-11
  \item \textbf{Test Coverage:} 100\%
  \item \textbf{Known Vulnerable Platforms:} Felix;
	Equinox;
	Knopflerfish;
	Concierge;
	SFelix
\end{itemize*}

%% file: vulnerabilityPatterns/runtime.halt.tex
\subsubsection{Runtime.halt}

\paragraph{Vulnerability Reference}

\begin{itemize*}
  \item \textbf{Vulnerability Name:} Runtime.halt
  \item \textbf{Identifier:} Mb.java.2
  \item \textbf{Origin:} Ares research project `malicious-bundle'
  \item \textbf{Location of Exploit Code:} Application Code - Java API
  \item \textbf{Source:} JVM - Runtime API (Runtime.halt method)
  \item \textbf{Target:} Platform
  \item \textbf{Consequence Type:} Unavailability
  \item \textbf{Introduction Time:} Development
  \item \textbf{Exploit Time:} Execution
\end{itemize*}

\paragraph{Vulnerability Description}

\begin{itemize*}
  \item \textbf{Description:} A bundle that stops the platform by calling 'Runtime.getRuntime.halt(0)'
  \item \textbf{Preconditions:} No SecurityManager, or RuntimePermission `exitVM'
  \item \textbf{Attack Process:} -
  \item \textbf{Consequence Description:} The shutdown hooks are by-passed
  \item \textbf{See Also:} System.exit, Exec.Kill, Recursive Thread Creation
\end{itemize*}

\paragraph{Protection}

\begin{itemize*}
  \item \textbf{Existing Mechanisms:} Java Permissions
  \item \textbf{Enforcement Point:} Platform startup
  \item \textbf{Potential Mechanisms:} Code static Analysis 
  \item \textbf{Attack Prevention:} -
  \item \textbf{Reaction:} Restart the platform
\end{itemize*}

\paragraph{Vulnerability Implementation}

\begin{itemize*}
  \item \textbf{Code Reference:} Fr.inria.ares.runtime\_halt-0.1.jar
  \item \textbf{OSGi Profile:} J2SE-1.5
  \item \textbf{Date:} 2006-08-11
  \item \textbf{Test Coverage:} 100\%
  \item \textbf{Known Vulnerable Platforms:} Felix;
	Equinox;
	Knopflerfish;
	Concierge;
	SFelix
\end{itemize*}

%% file: vulnerabilityPatterns/recursiveThreadCreation.tex
\subsubsection{Recursive Thread Creation}

\paragraph{Vulnerability Reference}

\begin{itemize*}
  \item \textbf{Vulnerability Name:} Recursive Thread Creation
  \item \textbf{Identifier:} Mb.java.3
  \item \textbf{Origin:} Ares research project `malicious-bundle'
  \item \textbf{Location of Exploit Code:} Application Code - Java API
  \item \textbf{Source:} JVM - APIs (Thread API);
        Application Code (No Algorithm Safety - Java)
  \item \textbf{Target:} Platform
  \item \textbf{Consequence Type:} Unavailability
  \item \textbf{Introduction Time:} Development
  \item \textbf{Exploit Time:} Execution
\end{itemize*}

\paragraph{Vulnerability Description}

\begin{itemize*}
  \item \textbf{Description:} The execution platform is brought to crash by the creation of an exponential number of threads
  \item \textbf{Preconditions:} -
  \item \textbf{Attack Process:} Each thread created by the attack bundle creates three other threads, and contains a relatively small payload (a pdf file). An excessive number of StackOverflowErrors causes an OutOfMemoryError
  \item \textbf{Consequence Description:} -
  \item \textbf{See Also:} System.exit, Runtime.halt, Exec.kill, Exponential Object Creation
\end{itemize*}

\paragraph{Protection}

\begin{itemize*}
  \item \textbf{Existing Mechanisms:} -
  \item \textbf{Enforcement Point:} -
  \item \textbf{Potential Mechanisms:} - 
  \item \textbf{Attack Prevention:} Stop the ill-behaving thread
  \item \textbf{Reaction:} Restart the platform
\end{itemize*}

\paragraph{Vulnerability Implementation}

\begin{itemize*}
  \item \textbf{Code Reference:} Fr.inria.ares.exponentialthreadnumber-0.1.jar
  \item \textbf{OSGi Profile:} J2SE-1.5
  \item \textbf{Date:} 2006-08-21
  \item \textbf{Test Coverage:} 50\%
  \item \textbf{Known Vulnerable Platforms:} Felix;
	Equinox;
	Knopflerfish;
	Concierge;
	SFelix
\end{itemize*}

%% file: vulnerabilityPatterns/hangingThread.tex
\subsubsection{Hanging Thread}

\paragraph{Vulnerability Reference}

\begin{itemize*}
  \item \textbf{Vulnerability Name:} Hanging Thread
  \item \textbf{Identifier:} Mb.java.4
  \item \textbf{Origin:} Java puzzlers 77, 85 \cite{bloch05puzzlers}
  \item \textbf{Location of Exploit Code:} Application Code - Java API
  \item \textbf{Source:} JVM - APIs (Thread API);
        Application Code (Value of Method Parameters)
  \item \textbf{Target:} OSGi Element - Service;
	OSGi Element - Package
  \item \textbf{Consequence Type:} Unavailability
  \item \textbf{Introduction Time:} Development
  \item \textbf{Exploit Time:} Execution
\end{itemize*}

\paragraph{Vulnerability Description}

\begin{itemize*}
  \item \textbf{Description:} Thread that make the calling entity hang (service, or package)
  \item \textbf{Preconditions:} -
  \item \textbf{Attack Process:} Use the Thread.sleep call with a large sleep duration to make the execution hang
  \item \textbf{Consequence Description:} If the sleep call is performed in a synchronized block, the SIG\_KILL (Ctrl+C) signal is caught by the platform
  \item \textbf{See Also:} Infinite Startup Loop
\end{itemize*}

\paragraph{Protection}

\begin{itemize*}
  \item \textbf{Existing Mechanisms:} -
  \item \textbf{Enforcement Point:} -
  \item \textbf{Potential Mechanisms:} Code static Analysis 
  \item \textbf{Attack Prevention:} -
  \item \textbf{Reaction:} -
\end{itemize*}

\paragraph{Vulnerability Implementation}

\begin{itemize*}
  \item \textbf{Code Reference:} Fr.inria.ares.hangingthread-0.1.jar, fr.inria.ares.hangingthread2-0.1.jar
  \item \textbf{OSGi Profile:} J2SE-1.5
  \item \textbf{Date:} 2006-08-28
  \item \textbf{Test Coverage:} 20\%
  \item \textbf{Known Vulnerable Platforms:} Felix;
	Equinox;
	Knopflerfish;
	Concierge;
	SFelix
\end{itemize*}

%% file: vulnerabilityPatterns/sleepingBundle.tex
\subsubsection{Sleeping Bundle}

\paragraph{Vulnerability Reference}

\begin{itemize*}
  \item \textbf{Vulnerability Name:} Sleeping Bundle
  \item \textbf{Identifier:} Mb.java.5
  \item \textbf{Origin:} Ares research project `malicious-bundle'
  \item \textbf{Location of Exploit Code:} Application Code - Java API
  \item \textbf{Source:} JVM - APIs (Thread API)
  \item \textbf{Target:} OSGi Element - Service;
	OSGi Element - Package
  \item \textbf{Consequence Type:} Performance Breakdown
  \item \textbf{Introduction Time:} Development
  \item \textbf{Exploit Time:} Execution
\end{itemize*}

\paragraph{Vulnerability Description}

\begin{itemize*}
  \item \textbf{Description:} A malicious bundle that goes to sleep during a specified amount of time before having finished its job (experience time is 50 sec.)
  \item \textbf{Preconditions:} -
  \item \textbf{Attack Process:} -
  \item \textbf{Consequence Description:} -
  \item \textbf{See Also:} Hanging Thread
\end{itemize*}

\paragraph{Protection}

\begin{itemize*}
  \item \textbf{Existing Mechanisms:} -
  \item \textbf{Enforcement Point:} -
  \item \textbf{Potential Mechanisms:} Code static Analysis 
  \item \textbf{Attack Prevention:} -
  \item \textbf{Reaction:} -
\end{itemize*}

\paragraph{Vulnerability Implementation}

\begin{itemize*}
  \item \textbf{Code Reference:} Fr.inria.ares.sleepingbundle-0.1.jar
  \item \textbf{OSGi Profile:} J2SE-1.5
  \item \textbf{Date:} 2006-08-28
  \item \textbf{Test Coverage:} 100\%
  \item \textbf{Known Vulnerable Platforms:} Felix;
	Equinox;
	Knopflerfish;
	Concierge;
	SFelix
\end{itemize*}

%% file: vulnerabilityPatterns/bigFileCreator.tex
\subsubsection{Big File Creator}

\paragraph{Vulnerability Reference}

\begin{itemize*}
  \item \textbf{Vulnerability Name:} Big File Creator
  \item \textbf{Identifier:} Mb.java.6
  \item \textbf{Origin:} Ares research project `malicious-bundle'
  \item \textbf{Location of Exploit Code:} Application Code - Java API
  \item \textbf{Source:} JVM - APIs (File API)
  \item \textbf{Target:} Platform
  \item \textbf{Consequence Type:} Performance Breakdown
  \item \textbf{Introduction Time:} Development
  \item \textbf{Exploit Time:} Execution
\end{itemize*}

\paragraph{Vulnerability Description}

\begin{itemize*}
  \item \textbf{Description:} A malicious bundle that create a big (relative to available resources) files to consume disk memory space
  \item \textbf{Preconditions:} No SecurityManager, or FilePermission `write' to the malicious bundl
  \item \textbf{Attack Process:} -
  \item \textbf{Consequence Description:} -
  \item \textbf{See Also:} Big Bundle Installer
\end{itemize*}

\paragraph{Protection}

\begin{itemize*}
  \item \textbf{Existing Mechanisms:} Java Permissions
  \item \textbf{Enforcement Point:} Platform startup
  \item \textbf{Potential Mechanisms:} Resource Control and Isolation - Disk Space (per user/bundle);
	Access Control - FileSystem (Limit the access to the FileSystem to the data directory of the bundle; control the size of the data created through the BundleContext)
  \item \textbf{Attack Prevention:} -
  \item \textbf{Reaction:} Erase files
\end{itemize*}

\paragraph{Vulnerability Implementation}

\begin{itemize*}
  \item \textbf{Code Reference:} Fr.inria.ares.bigfilecreator-0.1.jar
  \item \textbf{OSGi Profile:} J2SE-1.5
  \item \textbf{Date:} 2006-08-29
  \item \textbf{Test Coverage:} 00\%
  \item \textbf{Known Vulnerable Platforms:} Felix;
	Equinox;
	Knopflerfish;
	Concierge;
	SFelix
\end{itemize*}

%% file: vulnerabilityPatterns/codeObserver.tex
\subsubsection{Code Observer}

\paragraph{Vulnerability Reference}

\begin{itemize*}
  \item \textbf{Vulnerability Name:} Code Observer
  \item \textbf{Identifier:} Mb.java.7
  \item \textbf{Origin:} Ares research project `malicious-bundle'
  \item \textbf{Location of Exploit Code:} Application Code - Java API
  \item \textbf{Source:} JVM - APIs (Reflection API; ClassLoader API)
  \item \textbf{Target:} OSGi Element - Package;
	OSGi Element - Service
  \item \textbf{Consequence Type:} Undue Access
  \item \textbf{Introduction Time:} Development
  \item \textbf{Exploit Time:} Execution
\end{itemize*}

\paragraph{Vulnerability Description}

\begin{itemize*}
  \item \textbf{Description:} A component that observes the content of another one
  \item \textbf{Preconditions:} No SecurityManager, or ReflectPermission
  \item \textbf{Attack Process:} Use of the reflection API and the ClassLoader API
  \item \textbf{Consequence Description:} Observation of the implementation of the published packages and services, the classes that are aggregated to these packages and services, and classes which name are known (or guessed)
  \item \textbf{See Also:} Component Data Modifier, Hidden Method Launcher
\end{itemize*}

\paragraph{Protection}

\begin{itemize*}
  \item \textbf{Existing Mechanisms:} Java Permissions
  \item \textbf{Enforcement Point:} Platform startup
  \item \textbf{Potential Mechanisms:} - 
  \item \textbf{Attack Prevention:} -
  \item \textbf{Reaction:} -
\end{itemize*}

\paragraph{Vulnerability Implementation}

\begin{itemize*}
  \item \textbf{Code Reference:} Privateclassspy/fr.inria.ares.serviceabuser-0.1.jar
  \item \textbf{OSGi Profile:} J2SE-1.5
  \item \textbf{Date:} 2007-02-12
  \item \textbf{Test Coverage:} 100\%
  \item \textbf{Known Vulnerable Platforms:} Felix;
	Equinox;
	Knopflerfish;
	SFelix
\end{itemize*}

%% file: vulnerabilityPatterns/componentDataModifier.tex
\subsubsection{Component Data Modifier}

\paragraph{Vulnerability Reference}

\begin{itemize*}
  \item \textbf{Vulnerability Name:} Component Data Modifier
  \item \textbf{Extends:} Code Observer
  \item \textbf{Identifier:} Mb.java.8
  \item \textbf{Origin:} Ares research project `malicious-bundle'
  \item \textbf{Location of Exploit Code:} Application Code - Java API
  \item \textbf{Source:} JVM - APIs (Reflection API; ClassLoader API)
  \item \textbf{Target:} OSGi Element - Package;
	OSGi Element - Service
  \item \textbf{Consequence Type:} Undue Access
  \item \textbf{Introduction Time:} Development
  \item \textbf{Exploit Time:} Execution
\end{itemize*}

\paragraph{Vulnerability Description}

\begin{itemize*}
  \item \textbf{Description:} A bundle that modifies the data (i.e. the value of the attributes of the classes) of another one
  \item \textbf{Preconditions:} No SecurityManager or ReflectPermission set; the name of the non-exported modified class must be known beforehand, and contain a public static field.
  \item \textbf{Attack Process:} Use of the reflection API and the ClassLoader API to access and modify the value of attributes
  \item \textbf{Consequence Description:} Modification of the public fields of the objects that are accessible from another bundle (service implementations, or objects that are attributes of these service implementations, or objects that are attributes of these latter objects, ...), or of the public static (non final) fields of classes which name is known
  \item \textbf{See Also:} Code Observer, Hidden Method Launcher
\end{itemize*}

\paragraph{Protection}

\begin{itemize*}
  \item \textbf{Existing Mechanisms:} Java Permissions
  \item \textbf{Enforcement Point:} Platform startup
  \item \textbf{Potential Mechanisms:} - 
  \item \textbf{Attack Prevention:} -
  \item \textbf{Reaction:} -
\end{itemize*}

\paragraph{Vulnerability Implementation}

\begin{itemize*}
  \item \textbf{Code Reference:} Privateclassmanipulator/fr.inria.ares.serviceabuser-0.1.jar
  \item \textbf{OSGi Profile:} J2SE-1.5
  \item \textbf{Date:} 2007-02-12
  \item \textbf{Test Coverage:} 100\%
  \item \textbf{Known Vulnerable Platforms:} Felix;
	Equinox;
	Knopflerfish;
	Concierge;
	SFelix
\end{itemize*}

%% file: vulnerabilityPatterns/hiddenMethodLauncher.tex
\subsubsection{Hidden Method Launcher}

\paragraph{Vulnerability Reference}

\begin{itemize*}
  \item \textbf{Vulnerability Name:} Hidden Method Launcher
  \item \textbf{Extends:} Code Observer
  \item \textbf{Identifier:} Mb.java.9
  \item \textbf{Origin:} Ares research project `malicious-bundle'
  \item \textbf{Location of Exploit Code:} Application Code - Java API
  \item \textbf{Source:} JVM - APIs (Reflection API; ClassLoader API)
  \item \textbf{Target:} OSGi Element - Package
  \item \textbf{Consequence Type:} Undue Access
  \item \textbf{Introduction Time:} Development
  \item \textbf{Exploit Time:} Execution
\end{itemize*}

\paragraph{Vulnerability Description}

\begin{itemize*}
  \item \textbf{Description:} A bundle that executes methods from classes that are not exported of provided as service. All classes that are referenced (directly or indirectly) as class attributes can be accessed. Only public methods can be invoked
  \item \textbf{Preconditions:} No SecurityManager, or ReflectPermission set
  \item \textbf{Attack Process:} Use of the reflection API and the ClassLoader API to access and executes methods in classes that are not exported by the bundle
  \item \textbf{Consequence Description:} -
  \item \textbf{See Also:} Code Observer, Component Data Modifier
\end{itemize*}

\paragraph{Protection}

\begin{itemize*}
  \item \textbf{Existing Mechanisms:} Java Permissions
  \item \textbf{Enforcement Point:} Platform startup
  \item \textbf{Potential Mechanisms:} - 
  \item \textbf{Attack Prevention:} -
  \item \textbf{Reaction:} -
\end{itemize*}

\paragraph{Vulnerability Implementation}

\begin{itemize*}
  \item \textbf{Code Reference:} Hiddenclassexecutor/fr.inria.ares.serviceabuser-0.1.jar
  \item \textbf{OSGi Profile:} J2SE-1.5
  \item \textbf{Date:} 2007-02-12
  \item \textbf{Test Coverage:} 100\%
  \item \textbf{Known Vulnerable Platforms:} Felix;
	Equinox;
	Knopflerfish;
	Concierge;
	SFelix
\end{itemize*}

%% file: vulnerabilityPatterns/memoryLoadInjection.tex
\subsubsection{Memory Load Injection}

\paragraph{Vulnerability Reference}

\begin{itemize*}
  \item \textbf{Vulnerability Name:} Memory Load Injection
  \item \textbf{Identifier:} Mb.java.10
  \item \textbf{Origin:} MOSGI Ares Research Project (OSGi Platform Monitoring)
  \item \textbf{Location of Exploit Code:} Application Code - Java API
  \item \textbf{Source:} Application Code (No Algorithm Safety - Java)
  \item \textbf{Target:} Platform
  \item \textbf{Consequence Type:} Performance Breakdown
  \item \textbf{Introduction Time:} Development
  \item \textbf{Exploit Time:} Execution
\end{itemize*}

\paragraph{Vulnerability Description}

\begin{itemize*}
  \item \textbf{Description:} A malicious bundle that consumes most of available memory (61,65 MB in the example)
  \item \textbf{Preconditions:} -
  \item \textbf{Attack Process:} Store a huge amount of data in a byte array
  \item \textbf{Consequence Description:} Only a limited memory space is available for the execution of programs
  \item \textbf{See Also:} Ramping Memory Load Injection, CPU Load Injection
\end{itemize*}

\paragraph{Protection}

\begin{itemize*}
  \item \textbf{Existing Mechanisms:} -
  \item \textbf{Enforcement Point:} -
  \item \textbf{Potential Mechanisms:} Code static Analysis ;
	Resource Control and Isolation - Memory 
  \item \textbf{Attack Prevention:} -
  \item \textbf{Reaction:} -
\end{itemize*}

\paragraph{Vulnerability Implementation}

\begin{itemize*}
  \item \textbf{Code Reference:} Fr.inria.ares.memloadinjector-0.1.jar
  \item \textbf{OSGi Profile:} J2SE-1.5
  \item \textbf{Date:} 2006-08-24
  \item \textbf{Test Coverage:} 100\%
  \item \textbf{Known Vulnerable Platforms:} Felix;
	Equinox;
	Knopflerfish;
	Concierge;
	SFelix
\end{itemize*}

%% file: vulnerabilityPatterns/standAloneInfiniteLoop.tex
\subsubsection{Stand Alone Infinite Loop}

\paragraph{Vulnerability Reference}

\begin{itemize*}
  \item \textbf{Vulnerability Name:} Stand Alone Infinite Loop
  \item \textbf{Identifier:} Mb.java.11
  \item \textbf{Origin:} Ares research project `malicious-bundle'
  \item \textbf{Location of Exploit Code:} Application Code - Java Code
  \item \textbf{Source:} Application Code (No Algorithm Safety - Java)
  \item \textbf{Target:} Platform
  \item \textbf{Consequence Type:} Performance Breakdown
  \item \textbf{Introduction Time:} Development
  \item \textbf{Exploit Time:} Execution
\end{itemize*}

\paragraph{Vulnerability Description}

\begin{itemize*}
  \item \textbf{Description:} A void loop in a lonesome thread that consumes much of the available CPU
  \item \textbf{Preconditions:} -
  \item \textbf{Attack Process:} Infinite loop launched in an independent thread
  \item \textbf{Consequence Description:} -
  \item \textbf{See Also:} Infinite Startup Loop, CPU Load Injection
\end{itemize*}

\paragraph{Protection}

\begin{itemize*}
  \item \textbf{Existing Mechanisms:} -
  \item \textbf{Enforcement Point:} -
  \item \textbf{Potential Mechanisms:} Code static Analysis ;
	Resource Control and Isolation - CPU 
  \item \textbf{Attack Prevention:} -
  \item \textbf{Reaction:} -
\end{itemize*}

\paragraph{Vulnerability Implementation}

\begin{itemize*}
  \item \textbf{Code Reference:} Fr.inria.ares.standaloneloop-0.1.jar
  \item \textbf{OSGi Profile:} J2SE-1.5
  \item \textbf{Date:} 2006-09-22
  \item \textbf{Test Coverage:} 10\%
  \item \textbf{Known Vulnerable Platforms:} Felix;
	Equinox;
	Knopflerfish;
	Concierge;
	SFelix
\end{itemize*}

%% file: vulnerabilityPatterns/infiniteLoopInMethodCall.tex
\subsubsection{Infinite Loop in Method Call}

\paragraph{Vulnerability Reference}

\begin{itemize*}
  \item \textbf{Vulnerability Name:} Infinite Loop in Method Call
  \item \textbf{Identifier:} Mb.java.12
  \item \textbf{Origin:} Java puzzlers 26 to 33 \cite{bloch05puzzlers}
  \item \textbf{Location of Exploit Code:} Application Code - Java Code
  \item \textbf{Source:} Application Code (No Algorithm Safety - Java)
  \item \textbf{Target:} Platform;
	OSGi Element - Service;
	OSGi Element - Package
  \item \textbf{Consequence Type:} Performance Breakdown - Platform;
	Unavailability - Service, Package
  \item \textbf{Introduction Time:} Development
  \item \textbf{Exploit Time:} Execution
\end{itemize*}

\paragraph{Vulnerability Description}

\begin{itemize*}
  \item \textbf{Description:} An infinite loop is executed in a method call (at class use, package use)
  \item \textbf{Preconditions:} -
  \item \textbf{Attack Process:} An infinite loop is executed in a method call
  \item \textbf{Consequence Description:} Block the calling entity (the calling class or service. Because of the infinite loop, most CPU resource is consumed
  \item \textbf{See Also:} CPU Load Injection, Stand-alone Infinite Loop, Hanging Thread
\end{itemize*}

\paragraph{Protection}

\begin{itemize*}
  \item \textbf{Existing Mechanisms:} -
  \item \textbf{Enforcement Point:} -
  \item \textbf{Potential Mechanisms:} Code static Analysis ;
	Resource Control and Isolation - CPU 
  \item \textbf{Attack Prevention:} -
  \item \textbf{Reaction:} -
\end{itemize*}

\paragraph{Vulnerability Implementation}

\begin{itemize*}
  \item \textbf{Code Reference:} Fr.inria.ares.infiniteloopinmethodcall-0.1.jar
  \item \textbf{OSGi Profile:} J2SE-1.5
  \item \textbf{Date:} 2006-08-24
  \item \textbf{Test Coverage:} 10\%
  \item \textbf{Known Vulnerable Platforms:} Felix;
	Equinox;
	Knopflerfish;
	Concierge;
	SFelix
\end{itemize*}

%% file: vulnerabilityPatterns/exponentialObjectCreation.tex
\subsubsection{Exponential Object Creation}

\paragraph{Vulnerability Reference}

\begin{itemize*}
  \item \textbf{Vulnerability Name:} Exponential Object Creation
  \item \textbf{Identifier:} Mb.java.13
  \item \textbf{Origin:} Ares research project `malicious-bundle'
  \item \textbf{Location of Exploit Code:} Application Code - Java Code
  \item \textbf{Source:} Application Code (No Algorithm Safety - Java)
  \item \textbf{Target:} OSGi Element - Service;
	OSGi Element - Package
  \item \textbf{Consequence Type:} Unavailability
  \item \textbf{Introduction Time:} Development
  \item \textbf{Exploit Time:} Execution
\end{itemize*}

\paragraph{Vulnerability Description}

\begin{itemize*}
  \item \textbf{Description:} Objects are created in a exponential way
  \item \textbf{Preconditions:} -
  \item \textbf{Attack Process:} A given object create in its constructor 3 instances of object of the same class
  \item \textbf{Consequence Description:} The method call aborts with a 'StackOverflowError'
  \item \textbf{See Also:} Recursive Thread Creation
\end{itemize*}

\paragraph{Protection}

\begin{itemize*}
  \item \textbf{Existing Mechanisms:} -
  \item \textbf{Enforcement Point:} -
  \item \textbf{Potential Mechanisms:} Code static Analysis ;
	Resource Control and Isolation - Memory 
  \item \textbf{Attack Prevention:} -
  \item \textbf{Reaction:} -
\end{itemize*}

\paragraph{Vulnerability Implementation}

\begin{itemize*}
  \item \textbf{Code Reference:} Fr.inria.ares.exponentialobjectcreation-0.1
  \item \textbf{OSGi Profile:} J2SE-1.5
  \item \textbf{Date:} 2007-01-09
  \item \textbf{Test Coverage:} 50\%
  \item \textbf{Known Vulnerable Platforms:} Felix;
	Equinox;
	Knopflerfish;
	Concierge;
	SFelix
\end{itemize*}

%% file: vulnerabilityPatterns/launchHiddenBundle.tex
\subsubsection{Launch a Hidden Bundle}

\paragraph{Vulnerability Reference}

\begin{itemize*}
  \item \textbf{Vulnerability Name:} Launch a Hidden Bundle
  \item \textbf{Identifier:} Mb.osgi.6
  \item \textbf{Origin:} Ares research project `malicious-bundle'
  \item \textbf{Location of Exploit Code:} Application Code - OSGi API
  \item \textbf{Source:} OSGi Platform - Life-Cycle Layer (Bundle Management);
        JVM - APIs (File API)
  \item \textbf{Target:} Platform
  \item \textbf{Consequence Type:} Undue Access
  \item \textbf{Introduction Time:} Development
  \item \textbf{Exploit Time:} Execution
\end{itemize*}

\paragraph{Vulnerability Description}

\begin{itemize*}
  \item \textbf{Description:} A bundle that launches another bundle it contains (the contained bundle could be masqued as with a 'MyFile.java' file name)
  \item \textbf{Preconditions:} No SecurityManager, or OSGi PermissionAdmin and FilePermission `write' for the malicious bundle
  \item \textbf{Attack Process:} A bundle creates a new bundle on the file system, and launches i
  \item \textbf{Consequence Description:} A non foreseen bundle is installed. If install time checking exists (such as digital signature), it passes through the verification process
  \item \textbf{See Also:} Pirat Bundle Manager
\end{itemize*}

\paragraph{Protection}

\begin{itemize*}
  \item \textbf{Existing Mechanisms:} Java Permissions
  \item \textbf{Enforcement Point:} Platform startup
  \item \textbf{Potential Mechanisms:} - 
  \item \textbf{Attack Prevention:} -
  \item \textbf{Reaction:} Uninstall the malicious bundle
\end{itemize*}

\paragraph{Vulnerability Implementation}

\begin{itemize*}
  \item \textbf{Code Reference:} Fr.inria.ares.silentloader-0.1.jar, fr.inria.ares.silentloader.concierge-0.1.jar (without swing)
  \item \textbf{OSGi Profile:} J2SE-1.5
  \item \textbf{Date:} 2006-10-28
  \item \textbf{Test Coverage:} 100\%
  \item \textbf{Known Vulnerable Platforms:} Felix;
	Equinox;
	Knopflerfish;
	Concierge;
	SFelix
\end{itemize*}

%% file: vulnerabilityPatterns/piratBundleManager.tex
\subsubsection{Pirat Bundle Manager}

\paragraph{Vulnerability Reference}

\begin{itemize*}
  \item \textbf{Vulnerability Name:} Pirat Bundle Manager
  \item \textbf{Identifier:} Mb.osgi.7
  \item \textbf{Origin:} Ares research project `malicious-bundle'
  \item \textbf{Location of Exploit Code:} Application Code - OSGi API
  \item \textbf{Source:} OSGi Platform - Life-Cycle Layer (Bundle Management)
  \item \textbf{Target:} OSGi Element - Bundle
  \item \textbf{Consequence Type:} Undue Access
  \item \textbf{Introduction Time:} Development
  \item \textbf{Exploit Time:} Execution
\end{itemize*}

\paragraph{Vulnerability Description}

\begin{itemize*}
  \item \textbf{Description:} A bundle that manages others without being requested to do so(here: stops and starts the victim bundle)
  \item \textbf{Preconditions:} No SecurityManager, or OSGi Permission Admin
  \item \textbf{Attack Process:} The pirat bundle accesses to the bundle context, and then to its victim bundle
  \item \textbf{Consequence Description:} -
  \item \textbf{See Also:} Launch Hidden Bundle
\end{itemize*}

\paragraph{Protection}

\begin{itemize*}
  \item \textbf{Existing Mechanisms:} Java Permissions
  \item \textbf{Enforcement Point:} Platform startup
  \item \textbf{Potential Mechanisms:} - 
  \item \textbf{Attack Prevention:} -
  \item \textbf{Reaction:} -
\end{itemize*}

\paragraph{Vulnerability Implementation}

\begin{itemize*}
  \item \textbf{Code Reference:} Fr.inria.ares.piratbundlemanager-0.1.jar, fr.inria.ares.piratbundlemanager.concierge-0.1.jar (no swing)
  \item \textbf{OSGi Profile:} J2SE-1.5
  \item \textbf{Date:} 2006-10-30
  \item \textbf{Test Coverage:} 40\%
  \item \textbf{Known Vulnerable Platforms:} Felix;
	Equinox;
	Knopflerfish;
	Concierge;
	SFelix
\end{itemize*}

%% file: vulnerabilityPatterns/zombieData.tex
\subsubsection{Zombie Data}

\paragraph{Vulnerability Reference}

\begin{itemize*}
  \item \textbf{Vulnerability Name:} Zombie Data
  \item \textbf{Identifier:} Mb.osgi.8
  \item \textbf{Origin:} Ares research project `malicious-bundle'
  \item \textbf{Location of Exploit Code:} Application Code - OSGi API
  \item \textbf{Source:} OSGi Platform - Life-Cycle Layer (No Removal of Uninstalled Bundle Data)
  \item \textbf{Target:} Platform
  \item \textbf{Consequence Type:} Performance Breakdown
  \item \textbf{Introduction Time:} Development
  \item \textbf{Exploit Time:} Execution
\end{itemize*}

\paragraph{Vulnerability Description}

\begin{itemize*}
  \item \textbf{Description:} Data stored in the local OSGi data store are not deleted when the related bundle is uninstalled. It thus becomes unavailable and consumes disks space (especially on resource constraint devices)
  \item \textbf{Preconditions:} No SecuriyManager, or FilePermission set
  \item \textbf{Attack Process:} -
  \item \textbf{Consequence Description:} -
  \item \textbf{See Also:} Big File Creator
\end{itemize*}

\paragraph{Protection}

\begin{itemize*}
  \item \textbf{Existing Mechanisms:} Java Permissions
  \item \textbf{Enforcement Point:} Platform startup
  \item \textbf{Potential Mechanisms:} OSGi Platform Modification - Bundle Uninstall Process (Delete Bundle Data when Bundles are uninstalled)
  \item \textbf{Attack Prevention:} -
  \item \textbf{Reaction:} Erase files
\end{itemize*}

\paragraph{Vulnerability Implementation}

\begin{itemize*}
  \item \textbf{Code Reference:} Fr.inria.ares.bigfilecreator-0.1.jar
  \item \textbf{OSGi Profile:} J2SE-1.5
  \item \textbf{Date:} 2007-04-20
  \item \textbf{Test Coverage:} 100\%
  \item \textbf{Known Vulnerable Platforms:} Felix;
	Concierge
  \item \textbf{Known Robust Platforms:} Equinox;
        Knopflerfish;
        SFelix
\end{itemize*}

%% file: vulnerabilityPatterns/cycleBetweenServices.tex
\subsubsection{Cycle Between Services}

\paragraph{Vulnerability Reference}

\begin{itemize*}
  \item \textbf{Vulnerability Name:} Cycle Between Services
  \item \textbf{Identifier:} Mb.osgi.9
  \item \textbf{Origin:} Ares research project `malicious-bundle'
  \item \textbf{Location of Exploit Code:} Application Code - OSGi API
  \item \textbf{Source:} OSGi Platform - Service Layer (Architecture of the Application - No Validation of Service Dependency)
  \item \textbf{Target:} OSGi Element - Service;
	OSGi Element - Package
  \item \textbf{Consequence Type:} Unavailability
  \item \textbf{Introduction Time:} Service Publication or Resolution
  \item \textbf{Exploit Time:} Execution
\end{itemize*}

\paragraph{Vulnerability Description}

\begin{itemize*}
  \item \textbf{Description:} A cycle exists in the services call
  \item \textbf{Preconditions:} -
  \item \textbf{Attack Process:} Service 1 calls service 2, which calls service 1. The attack can be implemented as a fake 'service 2', which calls service 1 back instead of return properly from the method that service 1 called
  \item \textbf{Consequence Description:} `java.lang.StackOverflowError', service 1 can not be executed
  \item \textbf{See Also:} -
\end{itemize*}

\paragraph{Protection}

\begin{itemize*}
  \item \textbf{Existing Mechanisms:} -
  \item \textbf{Enforcement Point:} -
  \item \textbf{Potential Mechanisms:} Service-level dependency validation 
  \item \textbf{Attack Prevention:} -
  \item \textbf{Reaction:} -
\end{itemize*}

\paragraph{Vulnerability Implementation}

\begin{itemize*}
  \item \textbf{Code Reference:} Fr.inria.ares.clientserver1-0.1.jar, fr.inria.ares.clientserver2-0.1.jar
  \item \textbf{OSGi Profile:} J2SE-1.5
  \item \textbf{Date:} 2006-10-28
  \item \textbf{Test Coverage:} 100\%
  \item \textbf{Known Vulnerable Platforms:} Felix;
	Equinox;
	Knopflerfish;
	Concierge;
	SFelix
\end{itemize*}

%% file: vulnerabilityPatterns/numerousServiceRegistration.tex
\subsubsection{Numerous Service Registration}

\paragraph{Vulnerability Reference}

\begin{itemize*}
  \item \textbf{Vulnerability Name:} Numerous Service Registration
  \item \textbf{Identifier:} Mb.osgi.10
  \item \textbf{Origin:} Ares research project `malicious-bundle'
  \item \textbf{Location of Exploit Code:} Application Code - OSGi API
  \item \textbf{Source:} OSGi Platform - Service Layer (Uncontrolled Service Registration)
  \item \textbf{Target:} OSGi Element - Bundle;
	OSGi Element - Platform Management Utility
  \item \textbf{Consequence Type:} Performance Breakdown
  \item \textbf{Introduction Time:} Development
  \item \textbf{Exploit Time:} Execution
\end{itemize*}

\paragraph{Vulnerability Description}

\begin{itemize*}
  \item \textbf{Description:} Registration of a high number of (possibly identical) services through an loop
  \item \textbf{Preconditions:} No SecurityManager, or OSGi ServicePermission
  \item \textbf{Attack Process:} Publish a given service in a(n) (e.g. infinite) loop
  \item \textbf{Consequence Description:} Important duration of bundle stop
  \item \textbf{See Also:} -
\end{itemize*}

\paragraph{Protection}

\begin{itemize*}
  \item \textbf{Existing Mechanisms:} Java Permissions
  \item \textbf{Enforcement Point:} Platform startup
  \item \textbf{Potential Mechanisms:} OSGi Platform Modification - Service Publication (limitation of the number of services published in the framework)
  \item \textbf{Attack Prevention:} -
  \item \textbf{Reaction:} -
\end{itemize*}

\paragraph{Vulnerability Implementation}

\begin{itemize*}
  \item \textbf{Code Reference:} Fr.inria.ares.numerousservices-0.1.jar
  \item \textbf{OSGi Profile:} J2SE-1.5
  \item \textbf{Date:} 2007-01-09
  \item \textbf{Test Coverage:} 100\%
  \item \textbf{Known Vulnerable Platforms:} Felix;
	Equinox;
	Knopflerfish;
	Concierge
  \item \textbf{Known Robust Platforms:} SFelix
\end{itemize*}

%% file: vulnerabilityPatterns/freezingNumerousServiceRegistration.tex
\subsubsection{Freezing Numerous Service Registration}

\paragraph{Vulnerability Reference}

\begin{itemize*}
  \item \textbf{Vulnerability Name:} Freezing Numerous Service Registration
  \item \textbf{Identifier:} Mb.osgi.11
  \item \textbf{Origin:} Ares research project `malicious-bundle'
  \item \textbf{Location of Exploit Code:} Application Code - OSGi API
  \item \textbf{Source:} OSGi Platform - Service Layer (Uncontrolled Service Registration)
  \item \textbf{Target:} OSGi Element - Bundle;
	OSGi Element - Platform Management Utility
  \item \textbf{Consequence Type:} Performance Breakdown
  \item \textbf{Introduction Time:} Development
  \item \textbf{Exploit Time:} Execution
\end{itemize*}

\paragraph{Vulnerability Description}

\begin{itemize*}
  \item \textbf{Description:} Registration of a high number of (possibly identical) services through an loop, in the Concierge OSGi Platform implementation
  \item \textbf{Preconditions:} No SecurityManager, or OSGi ServicePermission; execution in the Concierge OSGi Platform
  \item \textbf{Attack Process:} Publish a given service in a(n) (e.g. infinite) loop
  \item \textbf{Consequence Description:} The Platform almost totally freeze. OutOfMemoryErrors are reported very frequently when the shell is used or when bundles perform actions.
  \item \textbf{See Also:} -
\end{itemize*}

\paragraph{Protection}

\begin{itemize*}
  \item \textbf{Existing Mechanisms:} Java Permissions
  \item \textbf{Enforcement Point:} Platform startup
  \item \textbf{Potential Mechanisms:} OSGi Platform Modification - Service Publication (limitation of the number of services published in the framework)
  \item \textbf{Attack Prevention:} -
  \item \textbf{Reaction:} -
\end{itemize*}

\paragraph{Vulnerability Implementation}

\begin{itemize*}
  \item \textbf{Code Reference:} Fr.inria.ares.numerousservices-0.1.jar
  \item \textbf{OSGi Profile:} J2SE-1.5
  \item \textbf{Date:} 2007-04-20
  \item \textbf{Test Coverage:} 100\%
  \item \textbf{Known Vulnerable Platforms:} Concierge
\end{itemize*}

%% file: vulnerabilityPatterns/executeHiddenClasses.tex
\subsubsection{Execute Hidden Classes}

\paragraph{Vulnerability Reference}

\begin{itemize*}
  \item \textbf{Vulnerability Name:} Execute Hidden Classes
  \item \textbf{Identifier:} Mb.osgi.12
  \item \textbf{Origin:} Ares research project `malicious-bundle'
  \item \textbf{Location of Exploit Code:} Bundle Fragment
  \item \textbf{Source:} OSGi Platform - Module Layer (Bundle Fragments)
  \item \textbf{Target:} OSGi Element - Package
  \item \textbf{Consequence Type:} Undue Access
  \item \textbf{Introduction Time:} Development
  \item \textbf{Exploit Time:} Execution
\end{itemize*}

\paragraph{Vulnerability Description}

\begin{itemize*}
  \item \textbf{Description:} A fragment bundle exports a pckage that is not made visible by the host. Other bundles can then execute the classes in this package
  \item \textbf{Preconditions:} No SecurityManager, or BundlePermission, `HOST' set to the host, and BundlePermission, `FRAGMENT', set to the malicious fragment
  \item \textbf{Attack Process:} -
  \item \textbf{Consequence Description:} Modification of static attributes, publication of hidden data or execution of secret procedure; Concierge does not support fragment, and does therefore not contains this vulnerability
  \item \textbf{See Also:} Fragment Substitution, Access Protected Package through split Packages
\end{itemize*}

\paragraph{Protection}

\begin{itemize*}
  \item \textbf{Existing Mechanisms:} Java Permissions
  \item \textbf{Enforcement Point:} Platform startup
  \item \textbf{Potential Mechanisms:} - 
  \item \textbf{Attack Prevention:} -
  \item \textbf{Reaction:} -
\end{itemize*}

\paragraph{Vulnerability Implementation}

\begin{itemize*}
  \item \textbf{Code Reference:} Usehiddenclass/packageexportfragment.jar\\+ usehiddenclass/fr.inria.ares.fragmentaccomplice-0.1.jar
  \item \textbf{OSGi Profile:} J2SE-1.5
  \item \textbf{Date:} 2007-02-14
  \item \textbf{Test Coverage:} 100\%
  \item \textbf{Known Vulnerable Platforms:} Felix;
	Equinox;
	Knopflerfish;
	SFelix
\end{itemize*}

%% file: vulnerabilityPatterns/fragmentSubstitution.tex
\subsubsection{Fragment Substitution}

\paragraph{Vulnerability Reference}

\begin{itemize*}
  \item \textbf{Vulnerability Name:} Fragment Substitution
  \item \textbf{Identifier:} Mb.osgi.13
  \item \textbf{Origin:} Ares research project `malicious-bundle'
  \item \textbf{Location of Exploit Code:} Bundle Fragment
  \item \textbf{Source:} OSGi Platform - Module Layer (Bundle Fragments)
  \item \textbf{Target:} OSGi Element - Bundle
  \item \textbf{Consequence Type:} Undue Access
  \item \textbf{Introduction Time:} Development
  \item \textbf{Exploit Time:} Execution
\end{itemize*}

\paragraph{Vulnerability Description}

\begin{itemize*}
  \item \textbf{Description:} A specific fragment bundle is replace by another, which provides the same classes but with malicious implementation
  \item \textbf{Preconditions:} No SecurityManager, or BundlePermission `HOST' and `FRAGMENT' set to the required bundles, and OSGi AdminPermission set to the substitutor bundle
  \item \textbf{Attack Process:} A malicious bundles uninstalls the current fragment, and install a malicious one (that is embedded in it) instead
  \item \textbf{Consequence Description:} The host bundle executes a false implementations of classes provided by a fragment; Concierge does not support fragment, and does therefore not contains this vulnerability
  \item \textbf{See Also:} Launch Hidden Bundle, Pirat Bundle Manager, Execute Hidden Class, Access Protected Package through split Packages
\end{itemize*}

\paragraph{Protection}

\begin{itemize*}
  \item \textbf{Existing Mechanisms:} Java Permissions
  \item \textbf{Enforcement Point:} Platform startup
  \item \textbf{Potential Mechanisms:} - 
  \item \textbf{Attack Prevention:} -
  \item \textbf{Reaction:} -
\end{itemize*}

\paragraph{Vulnerability Implementation}

\begin{itemize*}
  \item \textbf{Code Reference:} Fragmentprovidespackagestohost/fr.inria.ares.fragmentsubstitutor  (has an embedded  fragmentprovidespackagestohost/testfragmentclone bundle)
  \item \textbf{OSGi Profile:} J2SE-1.5
  \item \textbf{Date:} 2007-02-16
  \item \textbf{Test Coverage:} 100\%
  \item \textbf{Known Vulnerable Platforms:} Felix;
	Equinox;
	Knopflerfish;
	SFelix
\end{itemize*}

%% file: vulnerabilityPatterns/splitPackage.tex
\subsubsection{Access Protected Package through split Packages}

\paragraph{Vulnerability Reference}

\begin{itemize*}
  \item \textbf{Vulnerability Name:} Access Protected Package through split Packages
  \item \textbf{Identifier:} Mb.osgi.14
  \item \textbf{Origin:} Ares research project `malicious-bundle'
  \item \textbf{Location of Exploit Code:} Bundle Fragment
  \item \textbf{Source:} OSGi Platform - Module Layer (Bundle Fragments)
  \item \textbf{Target:} OSGi Element - Package
  \item \textbf{Consequence Type:} Undue Access
  \item \textbf{Introduction Time:} Development
  \item \textbf{Exploit Time:} Execution
\end{itemize*}

\paragraph{Vulnerability Description}

\begin{itemize*}
  \item \textbf{Description:} A package is built in the fragment, that have the same name than a package in the host. All package-protected classes and methods can then be accessed from the fragment, and through a proxy exported in the framework
  \item \textbf{Preconditions:} No SecurityManager, or BundlePermission `HOST' and `FRAGMENT' set to the suitable bundles
  \item \textbf{Attack Process:} -
  \item \textbf{Consequence Description:} Concierge does not support fragment, and does therefore not contains this vulnerability
  \item \textbf{See Also:} Execute Hidden Class, Fragment Substitution
\end{itemize*}

\paragraph{Protection}

\begin{itemize*}
  \item \textbf{Existing Mechanisms:} Java Permissions
  \item \textbf{Enforcement Point:} Platform startup
  \item \textbf{Potential Mechanisms:} Miscellaneous (the Java Archive defined 'seal' keyword in the Manifest File does not prevent the package to be completed by a split package in the fragment. It probably should)
  \item \textbf{Attack Prevention:} -
  \item \textbf{Reaction:} -
\end{itemize*}

\paragraph{Vulnerability Implementation}

\begin{itemize*}
  \item \textbf{Code Reference:} Fragmentsplitpackage/fr.inria.ares.testhostbundle-0.1.jar + \\fragmentsplitpackage/testfragment-0.1.jar + \\fragmentsplitpackage/fr.inria.ares.fragmentclient-0.1.jar
  \item \textbf{OSGi Profile:} J2SE-1.5
  \item \textbf{Date:} 2007-02-17
  \item \textbf{Test Coverage:} 100\%
  \item \textbf{Known Vulnerable Platforms:} Felix;
	Equinox;
	Knopflerfish;
	SFelix
\end{itemize*}

%% file: appendix/implementations.tex
\section{Attack Implementations}
\label{app:implementations}

The implementations of the presented attacks are given. Two types of attacks can be performed through several different implementations.

\subsection{Infinite Loops}

The various implementations of Infinites Loops in Java are given. They match the Vulnerability \textbf{mb.java.8} and \textbf{mb.java.12} in our catalog.

\subsubsection{First Option}

\setbox0\vbox{
    \begin{verbatim}
boolean condition==true;
While(condition);
      \end{verbatim}
}
\fbox{\box0}

\subsubsection{Second Option}

This implementation is given in the Java Puzzler \#26 \cite{bloch05puzzlers}.

\setbox0\vbox{
    \begin{verbatim}
public static final int END = Integer.MAX\_VALUE;
public static final int START = END - 100; // or any other start value
for (int i = START; i <= END; i++);
      \end{verbatim}
}
\fbox{\box0}

\subsubsection{Third Option}

This implementation is given in the Java Puzzler \#27 \cite{bloch05puzzlers}.

\setbox0\vbox{
    \begin{verbatim}
int i = 0;
while (-1 << i != 0)
{i++;}
      \end{verbatim}
}
\fbox{\box0}

\subsubsection{Fourth Option} 

This implementation is given in the Java Puzzler \#28 \cite{bloch05puzzlers}.

\setbox0\vbox{
    \begin{verbatim}
double i = 1.0/0.0; //can also be set to Double.POSITIVE\_INFINITY, 
//or a sufficiently big number (such as 1.0e17 or bigger)
while (i == i + 1);
      \end{verbatim}
}
\fbox{\box0}

\subsubsection{Fifth Option} 

This implementation is given in the Java Puzzler \#45 \cite{bloch05puzzlers}.

\setbox0\vbox{
    \begin{verbatim}
public static void main(String[] args) {
        workHard();
    }

    private static void workHard() {
        try {
            workHard();
        } finally {
            workHard();
        }
    }
      \end{verbatim}
}
\fbox{\box0}

\subsubsection{Other Implementations}

For further implementation of the infinite loop, you can also refer to
\begin{itemize*}
 \item the Java Puzzler \#29 (double i = Double.NaN;while(i! =i);), 
 \item the Java Puzzler \#30 (String i = ``a''; while (i != i + 0)), puzzler 31 ( short i = -1; while (i != 0) i >>>= 1;), 
 \item the Java Puzzler \#32 (Integer i = new Integer(0); Integer j = new Integer(0); while (i <= j \&\& j <= i \&\& i != j);), 
 \item the Java Puzzler \#33 (int i = Integer.MIN\_VALUE; while (i != 0 \&\& i == -i);(//or long i = Long.MIN\_VALUE), for() loop with unsuitable modification of the variables that intervene in the stop condition).
\end{itemize*}

\subsection{Hanging Thread}

The various implementations of Hanging Threads in Java are given. They match the Vulnerability \textbf{mb.java.9} in our catalog.

\subsubsection{First Option} 

This implementation is given in the Java Puzzler \#77 \cite{bloch05puzzlers}.

\setbox0\vbox{
    \begin{verbatim}
import java.util.Timer;
import java.util.TimerTask;

public class Stopper extends Thread{

private volatile boolean quittingTime = false;
    public void run() {
        while (!quittingTime)
            pretendToWork();
        System.out.println(``Beer is good'');
    }
    private void pretendToWork() {
        try {
            Thread.sleep(300); // Sleeping on the job?
        } catch (InterruptedException ex) { }
    }

    // It's quitting time, wait for worker - Called by good boss
    synchronized void quit() throws InterruptedException {
        quittingTime = true;
        join();
    }
    // Rescind quitting time - Called by evil boss
    synchronized void keepWorking() {
        quittingTime = false;
    }
        
        public void hang(){
                System.out.println(``HangingThread Stopper ready''
		+``to behave badly'');

                try
                {
                    final Stopper worker = new Stopper();
        worker.start();

        Timer t = new Timer(true); // Daemon thread
        t.schedule(new TimerTask() {
            public void run() { worker.keepWorking(); }
        }, 500);

        Thread.sleep(400);
        worker.quit();
                }
                catch( InterruptedException e)
                {
                    e.printStackTrace();
                }
        }
}
      \end{verbatim}
}
\fbox{\box0}

\subsubsection{Second Option} 

This implementation is given in the Java Puzzler \#85 \cite{bloch05puzzlers}.

\setbox0\vbox{
    \begin{verbatim}
static {
        Thread t = new Thread(new Runnable() {
            public void run() {
                initialized = true;
            }
        });
        t.start();
        try {
            t.join();
        } catch(InterruptedException e) {
            throw new AssertionError(e);
        }
    }
      \end{verbatim}
}
\fbox{\box0}

\subsubsection{Other Implementations}

For other implementations, see the JLint Manual\footnote{http://artho.com/jlint/manual.html} for `deadlock errors' (2 ocurrences).

%% file: appendix/xml2tex.tex
\section{XML2Tex Documentation Generator}
\label{xml2tex}

\begin{figure}[htb]
\centering
\epsfig{file=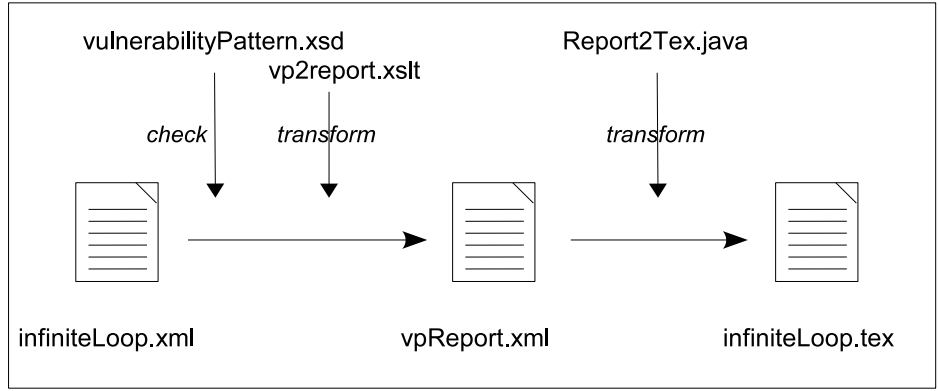, width=350pt}
\caption{XML2Tex Process}
\label{fig:xml2tex_process}
\end{figure}

To ease the validation of the Vulnerability Patterns and the generation of the catalog, we developed a small tool based on XML technologies, XML2Tex. The overall process of the XML2tex Documentation Generation process is given in Figure \ref{fig:xml2tex_process}.

Figure \ref{fig:xml2tex_process} presents the overview of the XML2tex Documentation Generation process.

First, the vulnerability pattern is checked against a reference XML Schema (XSD). Secondly, its is transformed into an XML report through XSL transformations. An XML Report is a specific XML file which is easily mappable to documentation: its contains in particular a title, paragraphs, and so on. Thirdly, the XML report is transformed into a Tex file through an Ad-Hoc parser named Report2Tex.

The validity of the Vulnerability Patterns of the catalog is guaranteed by their validation against the formal expression of the Pattern given in the appendix \ref{app:formalVulnerabilityPattern}. The well-formedness of the XML Report and of the Tex file are ensures by the successive parsers.

%% file: RR-6231.bbl
\newcommand{\etalchar}[1]{$^{#1}$}
\begin{thebibliography}{{OSG}05}

\bibitem[ACD{\etalchar{+}}75]{Abbott1975os}
R.~P. Abbott, J.~S. Chin, J.~E. Donnelley, W.~L. Konigsford, S.~Tokubo, and
  D.~A. Webb.
\newblock Security analysis and enhancements of computer operating systems.
\newblock Technical report, NATIONAL BUREAU OF STANDARDS WASHINGTONDC INST FOR
  COMPUTER SCIENCES AND TECHNOLOGY, December 1975.

\bibitem[AJB00]{avizienis00dependability}
A.Avizienis, J.C.Laprie, and B.Randell.
\newblock Fundamental concepts of dependability.
\newblock Technical Report No00493, LAAS (Toulouse, France), 2000.
\newblock 3rd Information Survivability Workshop (ISW'2000), Boston (USA),
  24-26 Octobre 2000, pp.7-12.

\bibitem[Ale77]{Alexander1977patternlanguage}
Christopher Alexander.
\newblock {\em A Pattern Language}.
\newblock Oxford University Press, 1977.

\bibitem[BAT06]{Bazaz2006contraints}
Anil Bazaz, James~D. Arthur, and Joseph~G. Tront.
\newblock Modeling security vulnerabilities: A constraints and assumptions
  perspective.
\newblock In {\em 2nd IEEE International Symposium on Dependable, Autonomic and
  Secure Computing (DASC'06)}, 2006.

\bibitem[BCHM99]{Baker1999cve}
David~W. Baker, Steven~M. Christey, William~H. Hill, and David~E. Mann.
\newblock The development of a common enumeration of vulnerabilities and
  exposures.
\newblock In {\em Second International Workshop on Recent Advances in Intrusion
  Detection}, 1999.

\bibitem[BG05]{bloch05puzzlers}
Joshua Bloch and Neal Gafter.
\newblock {\em Java Puzzlers - Traps, Pitfalls and Corner Cases}.
\newblock Pearson Education, June 2005.

\bibitem[BH78]{Bisbey1978protectionAnalysis}
Richard Bisbey and Dennis Hollingworth.
\newblock Protection analysis: Final report.
\newblock Technical Report ARPA ORDER NO. 2223, ISI/SR-78-13, Information
  Sciences Institute, University of Southern California, May 1978.

\bibitem[Blo01]{bloch01effectivejava}
Joshua Bloch.
\newblock {\em Effective Java Programming Language Guide}.
\newblock Addison-Wesley Professional, 2001.

\bibitem[Chr05]{Christey2005plover}
Steve Christey.
\newblock The preliminary list of vulnerability examples for researchers
  (plover).
\newblock In {\em NIST Workshop Defining the State of the Art of Software
  Security Tools, Gaithersburg, MD}, August 2005.

\bibitem[Chr06]{Christey2006interpretation}
Steven~M. Christey.
\newblock Open letter on the interpretation of "vulnerability statistics".
\newblock Bugtraq, Full-Disclosure Mailing list, January 2006.

\bibitem[CO05]{rfc4234}
D.~Crocker and P.~Overell.
\newblock Augmented bnf for syntax specifications: Abnf.
\newblock IETF RfC 4234, October 2005.

\bibitem[GHJV94]{gamma94pattern}
Erich Gamma, Richard Helm, Ralph Johnson, and John~M. Vlissides.
\newblock {\em Design Patterns: Elements of Reusable Object-Oriented Software}.
\newblock Addison-Wesley Professional Computing Series. Addison Wesley
  Professional., 1994.

\bibitem[GW05]{Gegick2005attackPatterns}
Michael Gegick and Laurie Williams.
\newblock Matching attack patterns to security vulnerabilities in
  software-intensive system designs.
\newblock {\em ACM SIGSOFT Software Engineering Notes}, 30(4), July 2005.

\bibitem[HL98]{howard98language}
John~D. Howard and Thomas~A. Longstaff.
\newblock A common language for computer security incidents.
\newblock Technical Report SAND98-8667, Sandia National Laboratories, USA,
  October 1998.

\bibitem[HLV05]{Howard2005_19sins}
Michael Howard, David LeBlanc, and John Viega.
\newblock {\em 19 Deadly Sins of Software Security}.
\newblock McGraw-Hill Osborne Media, July 2005.

\bibitem[Krs98]{Krsul1998softwareVulnerability}
Ivan~Victor Krsul.
\newblock {\em SOFTWARE VULNERABILITY ANALYSIS}.
\newblock PhD thesis, Purdue University, May 1998.

\bibitem[LBMC94]{landwehr94taxonomy}
Carl~E. Landwehr, Alan~R. Bull, John~P. McDermott, and William~S. Choi.
\newblock A taxonomy of computer program security flaws, with examples.
\newblock In {\em ACM Computing Surveys}, volume~26, pages 211--254, September
  1994.

\bibitem[LJ97]{Lindqvist1997classification}
Ulf Lindqvist and Erland Jonsson.
\newblock How to systematically classify computer security intrusions.
\newblock In {\em IEEE Symposium on Security and Privacy}, pages 154--163, May
  1997.

\bibitem[McG06]{mcgraw06softwaresecurity}
Gary McGraw.
\newblock {\em Software Security - Building Security In}.
\newblock Pearson Education, January 2006.

\bibitem[MCJ05]{martin2005enumeration}
Robert~A. Martin, Steven~M. Christey, and Joe Jarzombek.
\newblock The case for common flaw enumeration.
\newblock In {\em NIST Workshop on "Software Security Assurance Tools,
  Techniques, and Methods", Long Beach, CA., USA}, November 2005.

\bibitem[MEL01]{moore01modeling}
Andrew~P. Moore, Robert~J. Ellison, and Richard~C. Linger.
\newblock Attack modeling for information security and survivability.
\newblock Technical Report CMU/SEI-2001-TN-001, CMU/SEI, March 2001.

\bibitem[MM97]{Mowbray1997corbaDP}
Thomas~J. Mowbray and Raphael~C. Malveau.
\newblock {\em Corba Design Patterns}.
\newblock John Wiley \& Sons, January 1997.

\bibitem[Nec97]{necula97proofcarrying}
George~C. Necula.
\newblock Proof-carrying code.
\newblock In {\em Conference Record of {POPL}~'97: The 24th {ACM}
  {SIGPLAN}-{SIGACT} Symposium on Principles of Programming Languages}, pages
  106--119, Paris, France, jan 1997.

\bibitem[{OSG}05]{osgi05core}
{OSGI Alliance}.
\newblock Osgi service platform, core specification release 4.
\newblock Draft, 07 2005.

\bibitem[PF06]{parrend06deployment}
Pierre Parrend and Stephane Frenot.
\newblock Secure component deployment in the osgi(tm) release 4 platform.
\newblock Technical Report RT-0323, INRIA, June 2006.

\bibitem[PF07]{parrend2007sfelix}
Pierre Parrend and Stephane Frenot.
\newblock Supporting the secure deployment of osgi bundles.
\newblock In {\em First IEEE WoWMoM Workshop on Adaptive and DependAble
  Mission- and BUsiness-critical mobile Systems, Helsinki, Finland}, June 2007.

\bibitem[Sch03]{Schumacher2003}
Markus Schumacher.
\newblock {\em Security Engineering with Patterns}.
\newblock Springer Verlag, 2003.
\newblock LNCS n 2754.

\bibitem[SH05]{Seacord2005vulnerabilities}
Robert~C. Seacord and Allen Householder.
\newblock A structured approach to classifying security vulnerabilities.
\newblock Technical Report CMU/SEI-2005-TN-003, Carnegie Mellon University -
  Software Engineering Institute, January 2005.

\bibitem[SS73]{saltzer73}
Jerome~H. Saltzer and Michael~D. Schroeder.
\newblock The protection of information in computer systems.
\newblock In {\em Fourth ACM Symposium on Operating System Principles}, October
  1973.

\bibitem[{Sun}03]{jarSpec}
{Sun Microsystems, Inc.}
\newblock Jar file specification.
\newblock Sun Java Specifications, 2003.

\bibitem[TCM05]{tsipenyuk06taxonomy}
Katrina Tsipenyuk, Brian Chess, and Gary McGraw.
\newblock Seven pernicious kingdoms: A taxonomy of software security errors.
\newblock {\em IEEE Security \& Privacy}, 3(6):81--84, November/December 2005.

\bibitem[WKP05]{Weber2005softwareflaws}
Sam Weber, Paul~A. Karger, and Amit Paradkar.
\newblock A software flaw taxonomy: Aiming tools at security.
\newblock In {\em Software Engineering at Secure Systems - Building Trustworthy
  Applications}, June 2005.

\end{thebibliography}
